\documentclass[12pt]{article}

\newcommand{\biblist}{\begin{list}{}
{\listparindent 0.0cm \leftmargin 0.50cm \itemindent -0.50 cm
\labelwidth 0 cm \labelsep 0.50 cm
\usecounter{list}}\clubpenalty4000\widowpenalty4000}
\newcommand{\ebiblist}{\end{list}}

\usepackage{latexsym}
\usepackage{amsmath}
\usepackage{multirow}
\usepackage{amsfonts,amssymb}
\usepackage{graphicx,epsfig,color}
\usepackage{exscale}
\usepackage{amsthm}
\usepackage{calc}
\usepackage{flafter}
\usepackage{float}
\usepackage{dcolumn}
\usepackage{subfig} 
\usepackage{booktabs}
\usepackage{lscape}
\usepackage{bm}
\usepackage{hanging}

\usepackage[utf8]{inputenc}
\usepackage[english]{babel}

\newtheorem{example}{Example}[section]

\newcommand{\E}{\mathbb{E}}
\newcommand{\V}{\mathbb{V}}
\newcommand{\TAU}{\bm{\tau}}
\newcommand{\ALPHA}{\bm{\alpha}}
\newcommand{\BETA}{\bm{\beta}}
\newcommand{\ETA}{\bm{\eta}}

\begin{document}

\baselineskip .3in

\title{\bf {\huge  Efficient multiply robust imputation in the presence of influential units in surveys}}
\author{Sixia Chen\thanks{Department of Biostatistics and Epidemiology, University of Oklahoma Health Sciences Center, Oklahoma City, OK, U.S.A.}, ~ David Haziza\thanks{Department of Mathematics and Statistics, University of Ottawa, Ottawa, Canada;} ~ and ~ Victoire Michal\thanks{Department of Epidemiology, Biostatistics and Occupational Health, McGill University, Montreal, Canada.}}
\date{}
\maketitle

\begin{abstract}
Item nonresponse is a common issue in surveys. Because unadjusted estimators may be biased in the presence of nonresponse, it is common practice to impute the missing values with the objective of reducing the nonresponse bias as much as possible. However, commonly used imputation procedures may lead to unstable estimators of population totals/means when influential units are present in the set of respondents. In this article, we  consider the class of multiply robust imputation procedures that provide some protection against the failure of underlying model assumptions. We develop an efficient version of multiply robust estimators based on the concept of conditional bias, a  measure of influence. We  present the results of a simulation study to show the benefits of the proposed method in terms of bias and efficiency.
\end{abstract}


\section{Introduction}
Item nonresponse is ubiquitous in surveys conducted by National Statistical Offices. Most often, it is treated by some form of single imputation, whereby a missing value is replaced by some plausible value constructed under certain assumptions. The customary imputation process starts with specifying an imputation model describing the relationship between the variable $y$ requiring imputation, and a set of fully observed variables, $\mathbf{v},$ available for both respondents and nonrespondents. Defining an imputation model involves the selection of an appropriate set of predictors and, in the case of parametric imputation, the specification of a functional linking $y$ to $\mathbf{v}.$ Once the missing data have been imputed, a population total is readily estimated by computing a weighted sum of observed and imputed values. The validity of imputed estimators requires the first moment of the imputation model, $\E(y\mid \mathbf{v}),$  to be correctly specified. Misspecification of the first moment may result in significant bias. To protect against the misspecification of the imputation model, one can have recourse to multiply robust imputation procedures, whereby the imputer specifies multiple imputation models and/or multiple nonresponse models, where a nonresponse model is a set of assumptions describing the relationship between the response indicators (equal to 1 if $y$ is observed and equal to 0, otherwise) to a set of fully observed variables. The rationale behind multiply robust imputation procedures is to construct a set of imputed values by combining all the information contained in these multiple models. A procedure is said to be multiply robust if the resulting estimator remains consistent if all but one of the specified models are incorrectly specified, which is a desirable feature. The reader is referred to Han and Wang (2013), Han (2014a), Han (2014b), Chan and Yam (2014), Chen and Haziza (2017), Chen and Haziza (2019) for a discussion of multiply robust procedures. Double robustness (e.g., Robins et al., 1994; Scharfstein et al, 2005; Haziza and Rao, 2006; Kim and Park, 2006; Kang, Schafer, 2008; Cao and al., 2009; Kim and Haziza, 2014)  can be viewed as a special case of multiple robustness.\\

While multiply robust imputation procedures provide some protection against the failure of underlying model assumptions, the resulting estimators are generally vulnerable to the presence of influential units in the sample. A unit is said to be influential if its inclusion or exclusion from the computation has a large impact on the resulting estimate. In the presence of influential units, imputed estimators are (asymptotically) unbiased if the first moment of the imputation model is correctly specified but they may exhibit a large variance. At this stage, it is useful to distinguish influential units from gross measurement errors. The latter are identified and corrected at the data-editing stage. In contrast,  an influential unit corresponds to a respondent who exhibits a value that is correctly recorded. An influential unit may thus represent other similar units in the set of nonrespondents or in the non-sampled part of the population. This type of units has been called representative outliers by Chambers (1986) and are the focus of the current article. The issue of influential units is common  in business surveys. On the one hand, the distribution of economic variables is typically highly skewed, which generates a conducive ground for the presence of influential units. On the other hand, an influential unit can arise when the measure of size recorded on the sampling frame and used to stratify the population is considerably smaller than the size  recorded on the field. This unit is then placed in a stratum with smaller units. As a result, it will generally exhibit a large $y$-value combined with a large weight, which makes it potentially harmful. These units are often referred to as stratum jumpers. \\

To quantify the influence of a unit, we use the concept of conditional bias that was first suggested by  Mu\~{n}oz-Pichardo et al. (1995) in the customary independent and identically distributed (iid) setup and adapted by Moreno-Rebollo et al. (1999) and Moreno-Rebollo et al. (2002) in the survey sampling setup. At this stage, it is worth pointing out that a unit is influential/not influential with respect to a given configuration. In the context of imputation for missing survey data, a configuration consists of (i) the variable $y$ and its distribution in the population; (ii) the finite population parameter of interest; (iii) the sampling design and the associated estimator; (iv) whether or not the unit is present in the sample; (v) whether or not the unit responded to item $y$; (vi) the imputation procedure used to fill in the missing values. A  unit may have a large influence with respect to a given configuration but may have no influence with respect to another configuration. \\

In the ideal set-up of 100\% response, Beaumont et al. (2013) constructed an efficient version of the Horvitz-Thompson estimator based on the concept of conditional bias. Results of several empirical investigations suggest that the estimator of Beaumont et al. (2013) outperforms the Horvitz-Thompson in terms of mean square error when influential units are present in the sample. This is achieved at the expense of introducing a bias. In the absence of influential units in the sample, the estimator of Beaumont et al. (2013) suffers from a very slight loss of efficiency with respect to the Horvitz-Thompson estimator, which is a desirable feature.  Favre-Martinoz et al. (2016) extended the approach of Beaumont et al. (2013) to the case of two-phase sampling designs  and weighting for unit nonresponse. Doubly robust imputation procedures in the presence of influential units were considered in Dongmo Jiongo (2015) who extended the results of Beaumont et al. (2013) with the conditional bias evaluated with respect to two inferential frameworks: the nonresponse model framework and the imputation model framework. This led to two efficient estimators, one for each framework. In this article, we consider the case of multiply robust imputation and evaluate the conditional bias using a framework different from the ones considered in Dongmo Jiongo (2015). Our approach leads a single estimator, which is attractive from an imputer's perspective. \\

The paper is organized as follows. In Section 2, we briefly describe the approach of Beaumont et al. (2013) in the ideal scenario of 100\% response. In Section 3, we define the conditional bias of a unit and extend the results of Section 2 to the case of multiply robust imputation procedures. In Section 4, we present a bootstrap procedure for estimating the conditional bias of a unit. A calibrated imputation procedure is described in Section 5. In Section 6, we present the results from three empirical investigations, assessing the proposed method in terms of bias and efficiency. Some final remarks are given in Section 7. Finally, some technical details are relegated to the Appendix.

\section{Efficient complete data estimation}

Consider a finite population $U=\{1, \dots, i, \dots, N\}$ of size $N$. We are interested in estimating the population total, $t_y=\sum_{i \in U}y_i,$ of a survey variable $y$. We select a sample $S,$ of size $n,$ according to a given sampling design $p(S).$ Let  $I_i$ the sample selection indicator attached to unit $i,$ such that $I_i=1$ if $i \in S$ and $I_i=0$, otherwise. The first and second-order inclusion probabilities are respectively given by $\pi_i=\mathbb{P}(I_i=1), \  i \in U,$ and $\pi_{ik}=\mathbb{P}(I_i=1, I_k=1),\ i \in U,\ k \in U, \ i \neq k.$\\

A complete data or prototype estimator of $t_y$ is the Horvitz-Thompson estimator (Horvitz and Thompson, 1952)
\begin{equation}
\widehat{t}_{y, HT}=\sum_{i \in S}w_iy_i,
\label{Tot_Est_HT}
\end{equation} 
where $w_i=\pi_i^{-1}$denotes the sampling weight attached to unit $i$.  The Horvitz-Thompson estimator is design-unbiased for $t_y$; that is,  $\E_p\left(\widehat{t}_{y, HT}\right)=t_y,$ where the subscript $p$ denotes the sampling design. Under mild regularity conditions, it is also design-consistent for $t_y$ in the sense that $\widehat{t}_{y, HT}-t_y=O_p(N/\sqrt{n})$; see, e.g., Breidt and Opsomer (2017). \\

In the presence of influential units in the sample, the Horvitz-Thompson estimator may be highly unstable. In the ideal situation of 100\% response, Beaumont et al. (2013) proposed an efficient version of $\widehat{t}_{y, HT}$ based on the concept of conditional bias  (Moreno-Rebollo et al., 1999; Beaumont et al., 2013). Let $\theta_N$ be a finite population parameter and  $\widehat{\theta}$ be an estimator of $\theta_N$. The conditional bias attached to the $i$th sample unit is defined as $$B_{1i}=\E_p\left(\widehat{\theta} \mid I_i=1\right)- \theta_N.$$ 
If $\theta_N=t_y$ and $\widehat{\theta} = \widehat{t}_{y, HT}$, it can be shown that
\begin{equation}
\label{BiaisCond_HT}
B_{1i}=\E_p\left(\widehat{t}_{y, HT} \mid I_i=1\right)-t_y=\sum_{k \in U}\frac{\Delta_{ik}}{\pi_i\pi_k}y_k,
\end{equation}
where $\Delta_{ik}=\pi_{ik}-\pi_i\pi_k.$ Because the conditional bias (\ref{BiaisCond_HT}) depends on the complete set of population values, $y_1, \ldots, y_N,$ it is generally unknown. A conditionally unbiased estimator of 
 $B_{1i}$ is given by
\begin{equation}
\widehat{B}_{1i}=\sum_{k \in S}\frac{\Delta_{ik}}{\pi_k\pi_{ik}}y_k.
\end{equation}
That is, $\E_p\left(\widehat{B}_{1i} \mid I_i=1\right)={B}_{1i}.$ \\

We consider an efficient version of $\widehat{t}_{y, HT}$ of the form 
 $$\widehat{t}_{y, HT}^* = \widehat{t}_{y, HT} +\Delta(c),$$
where $\Delta(c)$ is a random variable that depends on the cut-off value $c$. Beaumont et al. (2013) suggested to determine the value of $\Delta(c)$ that minimizes the maximum absolute estimated conditional bias of $\widehat{t}_{y, HT}^*$. This leads to
\begin{equation}\label{Rob_HT}
\widehat{t}_{y, HT}^* = \widehat{t}_{y, HT} - \frac{\widehat{B}_{\min}+\widehat{B}_{\max}}{2},
\end{equation} 
where $\widehat{B}_{\min}=\min_{i \in S}(\widehat{B}_{1i})$ and $\widehat{B}_{\max}=\max_{i \in S}(\widehat{B}_{1i})$. Beaumont et al. (2013) showed empirically that $\widehat{t}_{y, HT}^*$ can be significantly more efficient than $\widehat{t}_{y, HT}$ when influential units are present in the sample. This is achieved at the expense of introducing a bias given by 
$$\E_p\left(\widehat{t}_{y, HT}^*\right)- t_y=-\frac{1}{2}\E_p (\widehat{B}_{\min}+\widehat{B}_{\max}).$$
However, under mild regularity conditions, the estimator (\ref{Rob_HT}) is design-consistent for $t_y$. That is, $\widehat{t}_{y, HT}^*-t_y=O_p(N/\sqrt{n}),$ which is a desirable property.

\section{Efficient estimation in the presence of missing data}
In practice, the survey variable $y$ may be prone to missing values. Let $r_i$ be a response indicator attached to unit $i$ such that $r_i=1$ if $y_i$ is observed and $r_i=0$ if $y_i$ is missing. Let $S_r=\{i \in S: r_i=1\}$ and  $S_{nr}=\{i \in S: r_i=0\}$  denote the set of respondents and the set of nonrespondents  to the survey variable $y,$ respectively. Throughout the paper, we assume that the data are Missing At Random (Rubin, 1976):
 $$p_i=\mathbb{P}(r_i=1\mid y_i, \mathbf{v}_i)=\mathbb{P}(r_i=1 \mid \mathbf{v}_i),$$ where $\mathbf{v}$ denotes a vector of fully observed variables.
 The true model linking the survey variable $y$ to the set of fully observed variables $\mathbf{v}$  is given by
 \begin{eqnarray*}\label{imp:mod}
\E_m(y_{i} \mid \mathbf{v}_i) &=&m(\mathbf{v}_i;\boldsymbol{\beta}), \nonumber\\
\mathrm{Cov}_m(y_{i},y_{j}\mid \mathbf{v}_i, \mathbf{v}_j )&=&0, \quad i \ne j, \nonumber \\
\mathrm{\V}_m(y_{i}\mid \mathbf{v}_i)&=&\sigma^{2}.
 \end{eqnarray*}
 Although we assume equal variances, our results can be easily extended to the case of  unequal variances.\\

Following Han and Wang (2013) and Chen and Haziza (2017), we consider two classes of models: 
\begin{itemize}
	\item[(i)] The class $\mathcal{C}_1$ of $J$ nonresponse models, each of the nonresponse model being a set of assumptions about the unknown nonresponse mechanism: $$\mathcal{C}_1=\{p^{(j)}(\mathbf{v}^{(j)}, \bm{\alpha}^{(j)}): j=1, \dots, J\},$$ where $p^{(j)}(\cdot, \ALPHA^{(j)})$ is a predetermined functional associated with the $j$th nonresponse model,  $\ALPHA^{(j)}$ is a vector of unknown parameters  and  $\mathbf{v}^{(j)}$ denotes the set of predictors included in the $j$th nonresponse model.
	\item[(ii)] The class $\mathcal{C}_2$ of $L$ imputation models, each model being a set of assumptions about the conditional distribution of $y$ given $\mathbf{v}$:
	$$\mathcal{C}_2=\{m^{(\ell)} (\mathbf{v}^{(\ell)}, \bm{\beta}^{(\ell)}): \ell=1, \dots, L\},$$ where $m^{(\ell)}(\cdot, \BETA^{(\ell)})$ is a predetermined functional associated with the  $\ell$th imputation model, $\BETA^{(\ell)}$ is a vector of unknown parameters and  $\mathbf{v}^{(\ell)}$  denotes the set of predictors included in the $\ell$th imputation model.
\end{itemize}  
Overall, the imputer specifies $J+L$ models that will be used in the construction of  the imputed values.\\

To construct the imputed values, $y_i^*, \ i\in S_{nr},$ we proceed as follows:
\begin{itemize}
	\item[(1)] We start by estimating the parameters ${\ALPHA}^{(j)}, \ j=1, \dots, J$ and   $\BETA^{(\ell)}, \ \ell=1, \dots, L$ by solving the following estimating equations:
$$S_{\bm{\alpha}}^{(j)}(\bm{\alpha}^{(j)}) =\sum_{i \in S}\phi_{pi} \frac{r_i-p^{(j)}(\mathbf{v}_i^{(j)}, \bm{\alpha}^{(j)})}{p^{(j)}(\mathbf{v}_i^{(j)}, \bm{\alpha}^{(j)})\{1-p^{(j)}(\mathbf{v}_i^{(j)}, \bm{\alpha}^{(j)})\}}\frac{\partial p^{(j)}(\mathbf{v}_i^{(j)}, \bm{\alpha}^{(j)})}{\partial \bm{\alpha}^{(j)}} =\mathbf{0}
$$ 
and
 $$S_{\bm{\beta}}^{(\ell)} (\bm{\beta}^{(\ell)} ) =\sum_{i \in S_r}\phi_{mi}\{y_i-m^{(\ell)} (\mathbf{v}_i^{(\ell)}, \bm{\beta}^{(\ell)} )\}\frac{\partial m^{(\ell)} (\mathbf{v}_i^{(\ell)}, \bm{\beta}^{(\ell)} )}{\partial \bm{\beta}^{(\ell)} } = \mathbf{0},$$
respectively, where $\phi_{pi}$ and $\phi_{mi}$ are two coefficients associated with unit $i$. In practice, these coefficients are either set to 1 or to $w_i.$

\item[(2)]  For each unit $i \in S$, we form the following vectors of size $J$ and $L,$ respectively:
$$\widehat{\mathbf{U}}_{pi}=\left(p^{(1)}(\mathbf{v}_i^{(1)}, \widehat{\bm{\alpha}}^{(1)}), \dots, p^J(\mathbf{v}_i^{(J)}, \widehat{\bm{\alpha}}^{(J)})\right)^\top$$
and 
$$\widehat{\mathbf{U}}_{mi}=\left(m^{(1)}(\mathbf{v}_i^{(1)}, \widehat{\bm{\beta}}^{(1)}), \dots, m^L(\mathbf{v}_i^{(L)}, \widehat{\bm{\beta}}^{(L)})\right)^\top.$$
To compress the information contained in the vector $\widehat{\mathbf{U}}_{pi}$, we fit a linear regression model with the response indicator $r$ as the dependent variable and the vector $\widehat{\mathbf{U}}_{p}$ as the set of predictors. This leads to the $J$-vector of estimated coefficients
$$\widehat{\bm{\eta}}_p = \left(\sum_{i \in S}w_i \widehat{\mathbf{U}}_{pi}\widehat{\mathbf{U}}^\top_{pi}\right)^{-1}\sum_{i \in S} w_i \widehat{\mathbf{U}}_{pi}r_i.$$ 
To compress the information contained in the vector $\widehat{\mathbf{U}}_{mi}$, we fit a linear regression model based on the responding units, with the survey variable $y$ as the dependent variable and the vector $\widehat{\mathbf{U}}_{m}$ as the set of predictors. This leads to the $L$-vector of estimated coefficients
$$\widehat{\bm{\eta}}_m = \left(\sum_{i \in S_r}w_i \widehat{\mathbf{U}}_{mi}\widehat{\mathbf{U}}^\top_{mi}\right)^{-1}\sum_{i \in S_r} w_i \widehat{\mathbf{U}}_{mi}y_i.$$
Finally, for each unit $i \in S$, we obtain the following two standardized scores:
$$\widehat{p}_i = \widehat{\mathbf{U}}^\top_{pi}\frac{\widehat{\bm{\eta}}^2_p}{\widehat{\bm{\eta}}^\top_p\widehat{\bm{\eta}}_p} \quad \mbox{and} \quad \widehat{m}_i = \widehat{\mathbf{U}}^\top_{mi}\frac{\widehat{\bm{\eta}}^2_m}{\widehat{\bm{\eta}}^\top_m\widehat{\bm{\eta}}_m}.$$
Here, if $\mathbf{a}=(a_1, \dots, a_h)^\top$ is a $h$-vector, $\mathbf{a}^2$ denotes the vector of square coefficients $(a_1^2, \dots, a_h^2)^\top$. 
\item[(3)] The imputed values $y_i^*, i \in S_{nr},$ are obtained by fitting a weighted linear regression model with $y$ as the dependent variable and $\mathbf{h}=(1, \widehat{m})^{\top} $ as the vector of predictors. The regression weights are given by  $w_i\left(\widehat{p}_i^{-1}-1\right)$, $i \in S_r$. This leads to   
\begin{equation}\label{imp_val_MR}
y_i^*=\mathbf{h}_i^\top \widehat{\bm{\tau}}, \ i \in S_{nr},
\end{equation}
where 
$$\displaystyle\widehat{\bm{\tau}}=\left(\sum_{i \in S_r}w_i\left(\widehat{p}_i^{-1}-1\right)\mathbf{h}_i\mathbf{h}_i^\top \right)^{-1}\sum_{i \in S_r}w_i\left(\widehat{p}_i^{-1}-1\right)  \mathbf{h}_iy_i.$$
\end{itemize}
When $J=0$ and $L=1$, the imputation procedure (\ref{imp_val_MR})  reduces to an imputation based on a single imputation model. When $J=1$ and $L=1$, the imputation procedure (\ref{imp_val_MR}) corresponds to a doubly robust imputation procedure.\\

Based on the observed values, $y_i, i \in S_r,$ and the imputed values, $y_i^*, i \in S_{nr},$ we construct an imputed estimator of $t_y$: 
\begin{equation}
\widehat{t}_{MR}=\sum_{i \in S_r}w_iy_i+\sum_{i \in S_{nr}}w_i\mathbf{h}_i^\top \widehat{\TAU}.
\label{Tot_Est_MR}
\end{equation}
The estimator (\ref{Tot_Est_MR}) is multiply robust in the sense that it remains consistent for $t_y$ if all but one of the $J+L$ models are incorrectly specified. That is,  if at least one one the $J+L$ models is correctly specified, we have
$$\widehat{t}_{MR}/t_y \overset{P}{\longrightarrow} 1$$
as $n \longrightarrow \infty$ and $N\longrightarrow \infty$; see, e.g., Chen and Haziza (2017).\\

While the imputation procedure (\ref{imp_val_MR}) provides some protection against model misspecification, the resulting estimator (\ref{Tot_Est_MR})  may be highly unstable in the presence of influential units. In this section, we develop an efficient version of $\widehat{t}_{MR}$ by extending the results of Beaumont et al. (2013). \\

We start by defining the concept of conditional bias in the context of imputation for missing data. We identify three sources of randomness: the imputation model that generates the $N$-vector of population $y$-values,  $\mathbf{y}_U=(y_1, \ldots, y_N)^{\top}$; the sampling design that generates the $N$-vector of sample selection indicators, $\mathbf{I}_U=(I_1, \ldots, I_N)^{\top}$; and  the nonresponse mechanism that generates the $N$-vector of response indicators, $\mathbf{r}_U=(r_1, \ldots, r_N)^{\top}$. Different combinations of these distributions may be used to assess the conditional bias of a unit. In the sequel,  the conditional bias is evaluated with respect to the sampling design. That is, in addition to the sets of predictors included in the imputation and nonresponse models, the vectors $\mathbf{y}_U$ and $\mathbf{r}_U$ will be treated as fixed. The conditional bias associated with unit $i$ of $ \widehat{t}_{MR}$ is thus defined as
\begin{equation}\label{Biais_Cond_MR}
B_{1i}^{(MR)}=\E_p\left(\widehat{t}_{MR}-t_y \mid I_i=1\right).
\end{equation}
The expectation on the right hand-side of (\ref{Biais_Cond_MR}) is intractable as the estimator $\widehat{t}_{MR}$ is a complex function of the sample selection indicators $I_1, \ldots, I_N.$ Therefore, we rely on the first-order Taylor expansion, which leads to
\begin{equation}\label{Taylor_exp}
\widehat{t}_{MR} = \sum_{k \in S} w_k\psi_k+o_p\left(\frac{N}{\sqrt{n}}\right),
\end{equation} 
where
\begin{align}\label{psi}
\psi_k &=y_k-\left(1-\frac{r_k}{p_k^\bullet}\right)(y_k-\mathbf{h}_k^{\bullet \top}\TAU^\bullet)\nonumber \\
&+ \sum_{j=1}^J\mathbf{A}_{\boldsymbol \alpha}^{(j)\bullet}\frac{r_k-p^{(j)}(\mathbf{v}_k, {\ALPHA}^{(j)\bullet})}{p^{(j)}(\mathbf{v}_k, {\ALPHA}^{(j)\bullet})\left(1-p^{(j)}(\mathbf{v}_k, {\ALPHA}^{(j)\bullet})\right)}\frac{\partial p^{(j)}(\mathbf{v}_k, {\ALPHA}^{(j)\bullet})}{\partial \ALPHA^{(j)}}\nonumber\\
&+ \sum_{\ell=1}^L\mathbf{A}_{\boldsymbol \beta}^{(\ell)\bullet}r_k\left(y_k- m^{(\ell)}(\mathbf{v}_k, {\BETA}^{(\ell)\bullet})\right)\frac{\partial m^{(\ell)}(\mathbf{v}_k, {\BETA}^{(\ell)\bullet})}{\partial \BETA^{(\ell)}}\nonumber \\
&+ \mathbf{A}_{p}^{\bullet}(r_k-\mathbf{U}_{pk}^{\top})\mathbf{U}_{pk}+\mathbf{A}_{m}^{\bullet}r_k(y_k-\mathbf{U}_{mk}^{\top})\mathbf{U}_{mk}\nonumber\\
&+\mathbf{A}_{\TAU}^{\bullet}r_k\frac{1-p^{(j)}(\mathbf{v}_k, {\ALPHA}^{(j)\bullet})}{p^{(j)}(\mathbf{v}_k, {\ALPHA}^{(j)\bullet})}(y_k-\mathbf{h}_k^{\bullet \top}\TAU^{\bullet}),
\end{align}
with $\ALPHA^\bullet$, $\BETA^\bullet$, $\ETA_p^\bullet$, $\ETA_m^\bullet$ and $\TAU^\bullet$ denoting the probability limits of  $\widehat{\ALPHA}$, $\widehat{\BETA}$, $\widehat{\ETA}_p$, $\widehat{\ETA}_m$ and $\widehat{\TAU},$ respectively. The derivations leading to (\ref{psi}) are shown in the Appendix.\\

Using (\ref{Taylor_exp}) and ignoring the higher-order terms, we obtain the following approximation of the conditional bias attached to unit $i$:
\begin{align}
B_{1i}^{(MR)} &= \E_p\left(\widehat{t}_{MR}-t_y\mid I_i=1\right) \nonumber \\
&\simeq \E_p\left(\sum_{k \in S}w_k\psi_k - \sum_{k \in U}\psi_k + \sum_{k \in U}\psi_k -t_y\mid I_i=1\right) \nonumber \\
&= \sum_{k \in U} \frac{\Delta_{ik}}{\pi_i\pi_k}\psi_k + \sum_{k \in U}\psi_k-t_y. \label{B_Cond_MR_psity}
\end{align}

The conditional bias in (\ref{B_Cond_MR_psity}) is unknown as it involves population quantities. An estimator of (\ref{B_Cond_MR_psity}) is given by
\begin{equation}
\widehat{B}_{1i}^{(MR)}=\sum_{k \in S}\frac{\Delta_{ik}}{\pi_{ik}\pi_k}\widehat{\psi}_k,
\label{Biais_Cond_MR_Est}
\end{equation} 
where $\widehat{\psi}_k$ in (\ref{Biais_Cond_MR_Est}) is obtained from (\ref{psi}) by replacing each unknown quantity with a corresponding estimator. 
Note that the estimator $\widehat{B}_{1i}^{(MR)}$ in \eqref{Biais_Cond_MR_Est} does not involve the term $\sum_{k \in U}\psi_k-t_y$ on the right hand-side of \eqref{B_Cond_MR_psity}. Indeed, an estimator  of $t_y$ is given by $\widehat{t}_{MR},$ whereas an estimator of  $\sum_{k \in U}\psi_k$ is given by $\sum_{k \in S}w_k\psi_k$. From (\ref{Taylor_exp}), we have,  $\widehat{t}_{MR} - \sum_{k \in S}w_k\psi_k=o_p(N/\sqrt{n})$, and, as a result, is ignored. \\

\begin{example}
Consider the case of an imputation procedure based on a single model ($J=0$ and $L=1$) with $m (\mathbf{v}_i, \bm{\beta})= \mathbf{v}_i^{\top} \bm{\beta}$. The linearized variable $\psi_k$ in (\ref{psi}) reduces to
\begin{equation}\label{Cond_bias_MR_LR}
\psi_k=y_k+(r_ka_k-1)(y_k-\mathbf{v}_k^\top\BETA^\bullet),
\end{equation}
where
$$a_k=1+(\mathbf{t}_{\mathbf{v}}-\mathbf{t}_{\mathbf{v}_r})^\top\mathbf{T}_r^{-1}\mathbf{v}_k$$
with $\mathbf{t}_{\mathbf{v}}=\sum_{i \in U}\mathbf{v}_i,$  $\mathbf{t}_{\mathbf{v}_r}=\sum_{i \in U}r_i\mathbf{v}_i,$ and $\mathbf{T}_r=\sum_{i \in U}r_i\mathbf{v}_i\mathbf{v}_i^{\top}$.
The estimated linearized variable $\widehat{\psi}_k$ is obtained by estimating each unknown quantity in (\ref{Cond_bias_MR_LR})  with a suitable estimator. This leads to
$$\widehat{\psi}_k=y_k+(r_k\widehat{a}_k-1)(y_k-\mathbf{v}_k^\top\widehat{\BETA}_r),$$
where 
$$\widehat{a}_k=1+(\widehat{\mathbf{t}}_{\mathbf{v}, HT}-\widehat{\mathbf{t}}_{\mathbf{v}_r})^\top\widehat{\mathbf{T}}_r^{-1}\mathbf{v}_k$$
with
 $\widehat{\mathbf{t}}_{\mathbf{v}, HT}=\sum_{i \in S}w_i\mathbf{v}_i,$  $\widehat{\mathbf{t}}_{\mathbf{v}_r}=\sum_{i \in S}w_ir_i\mathbf{v}_i,$  $\widehat{\mathbf{T}}_r=\sum_{i \in S}w_ir_i\mathbf{v}_i\mathbf{v}_i^{\top}$ and
$$\widehat{\BETA}_r=\widehat{\mathbf{T}}_r^{-1}\sum_{i \in S}w_ir_i\mathbf{v}_iy_i. $$
It follows that the estimated conditional bias associated with unit $i$ of the multiply robust estimator $\widehat{t}_{MR}$ is given by
\begin{equation}
\widehat{B}_{1i}^{(MR)}=\sum_{k \in S}\frac{\Delta_{ik}}{\pi_{ik}\pi_k}y_k+\sum_{k \in S}\frac{\Delta_{ik}}{\pi_{ik}\pi_k}(r_k\widehat{a}_k-1)(y_k-\mathbf{v}_k^\top\widehat{\BETA}_r).
\label{Est_Cond_Bias_Lin_Reg}
\end{equation} 
The first term on the right hand-side of (\ref{Est_Cond_Bias_Lin_Reg}) corresponds to the influence of unit $i$ on the sampling error, whereas the second term represents the effect of nonresponse and imputation on the influence of unit $i$. From (\ref{Est_Cond_Bias_Lin_Reg}), a unit has a large influence if its complete data conditional bias is large and/or its residual $y_k-\mathbf{v}_k^\top\widehat{\BETA}_r)$ is large and/or if the $\widehat{a}_k$ is large (which may indicate that the unit has a high leverage). Therefore, our measure accounts for all the components of the configuration described in Section 1.
\end{example}

As in Section 2, we consider an efficient version of $\widehat{t}_{MR}$ of the form 
$$\widehat{t}_{MR}^* = \widehat{t}_{MR} +\Delta(c).$$
We determine the value of $\Delta(c)$ that minimizes the maximum absolute conditional bias of $\widehat{t}_{MR}^*$. This leads to
\begin{equation}
\label{Tot_Est_MR_Rob}
\widehat{t}_{MR}^*=\widehat{t}_{MR}-\frac{\widehat{B}_{\min}^{(MR)}+\widehat{B}_{\max}^{(MR)}}{2},
\end{equation} 
where  $\widehat{B}_{\min}^{(MR)}=\min_{i \in S}(\widehat{B}_{1i}^{(MR)})$ and $\widehat{B}_{\max}^{(MR)}=\max_{i \in S}(\widehat{B}_{1i}^{(MR)})$. If at least one of the $J+L$ models is correctly specified, the bias of  $\widehat{t}_{MR}$ is negligible.  As a result, the bias of $\widehat{t}_{MR}^*$ can be approximated by
$$\E(\widehat{t}_{MR}^*)-t_y \approx -\frac{1}{2}\E\left(\widehat{B}_{\min}^{(MR)}+\widehat{B}_{\max}^{(MR)}\right),$$
where the expectation $\E(.)$ is evaluated with respect to the joint distribution induced by the imputation model, the nonresponse mechanism and the sampling design.

\section{Pseudo-population bootstrap procedure for estimating the conditional bias}
In Section 3, we derived an approximation of the conditional bias based on a first-order Taylor expansion. However, the derivation involved relatively tedious algebra. In this section, we describe a pseudo-population bootstrap procedure for estimating the conditional bias; see Mashreghi et al. (2016) for a discussion of bootstrap procedures in finite population sampling. The idea behind pseudo-population bootstrap procedures is to create a pseudo-population from the original sample. Bootstrap samples are then selected from the pseudo-population using the same sampling design utilized to select the original samples.\\

A general pseudo-population bootstrap algorithm can be described as follows:

\begin{itemize}
	\item [(i)]	Repeat the pair $(y_i, \pi_i)$, $\left\lfloor \pi_i^{-1} \right\rfloor$ times for all $i$ in $S$ to create, $U^f$, the fixed part of the pseudo-population.
	
	\item [(ii)] To complete the pseudo-population, $U^*$, draw $U^{c*}$ from $\{(y_i, \pi_i)\}_{i \in S}$ using the original sampling design with inclusion probability $\pi_i^{-1} - \left\lfloor \pi_i^{-1} \right\rfloor$ for the $i$th pair, leading to $U^*=U^f~\cup~U^{c*}$.  Compute the bootstrap parameter $t_y^*$ on the resulting pseudo-population $U^*=\{(y_i^*, \pi_i^*)\}$.
	
	\item [(iii)] Take a bootstrap sample $S^*$ from $U^*$ using the same sampling design that led to $S$.
	
	\item [(iv)] Let $ S^*_R=\{i \in S^*: \ r_i^*=1\}$. Impute the bootstrap missing values in $S^* \setminus S^*_R$ by applying the same imputation method used for the original missing data.  Compute the estimator $\widehat t_{MR}^{*}$ based on observed and imputed values in the bootstrap sample $S^*$. 
	\item [(v)]		Repeat Steps 1 to 4 a large number of times, $M$,  to get $t_{y, 1}^{*}, \ldots, t_{y, M}^{*}$ and $\widehat {t}_{MR, 1}^{*}, \ldots, \widehat {t}_{MR, M}^{*}$.
	
	\item [(vi)]	Let $S_{b, i}$ be the set of bootstrap samples that contain unit $i$. A bootstrap estimator of $B_{1i}^{(MR)}$ in (\ref{Biais_Cond_MR}) is given by
	$$
	\widehat{B}_{1i}^{(*MR)} = M_i^{-1} \sum_{S^* \subset S_{b, i}}(\widehat {t}_{MR, m}^{*} -  t_{y, m}^{*}),
	$$
where $M_i$ denotes the cardinality of $S_{b, i}.$ 
\end{itemize}

\section{Calibrated imputation procedure}
The proposed method described in Section 3 consists of (i) imputing the missing values according to (\ref{imp_val_MR}), (ii) computing the multiply robust estimator $\widehat{t}_{MR}$ given by (\ref{Tot_Est_MR}) and (iii) obtaining the efficient version $\widehat{t}_{MR}^*$ given by (\ref{Tot_Est_MR_Rob}). In this section, we suggest implementing the proposed method through a calibrated imputation procedure, which may be attractive from a secondary analyst's point of view. The concept of calibrated imputation has been considered in Ren and Chambers (2003), Beaumont and Alavi (2004) and Beaumont (2005), among others. \\

The rationale behind calibrated imputation is to find final imputed values $y_{iF}^*, \ i \in S_{nr},$ as close as possible to the preliminary imputed values  $y_i^*$ given by (\ref{imp_val_MR}) subject to 
\begin{equation}\label{cal_con}
\widehat{t}_{MR, F}\equiv \sum_{i \in S}w_ir_iy_i+\sum_{i \in S}w_i(1-r_i)y_{iF}^*=\widehat{t}_{MR}-\frac{\widehat{B}_{\min}^{MR}+\widehat{B}_{\max}^{MR}}{2}.\end{equation}
More specifically, we seek final imputed values $y_{iF}^*,\ i \in S_{nr},$ that minimize
$$\sum_{i \in S}q_i^{-1}G(y_{iF}^*/y_i^*),$$
subject to (\ref{cal_con}), where $G(\cdot)$ is a pseudo-distance function and $q_i >0$ is a known coefficient attached to unit $i$. The pseudo-distance function $G(\cdot)$ must satisfy the following properties: (i) $G(y_{iF}^*/y_i^*) \ge 0$ and $G(1)=0$; (ii) $G(\cdot)$ is differentiable with respect to $y_{iF}^*$; (iii) the derivatives $g(u)=\partial G(u)/\partial u$ are continuous; (iv) $G(\cdot)$ is strictly convex; see Deville and Särndal (1992) for a description of commonly used functions $G(\cdot)$. Seeking final imputed values $y_{iF}^*,\ i \in S_{nr},$ close to the preliminary values $y_i^*$ is desirable as the latter ensure that the resulting imputed estimator $\widehat{t}_{MR}$ is a consistent estimator of $t_y$ if at least one of the $J+L$ models is correctly specified. \\

For instance, if we use the generalized chi-square distance,  we seek final imputed values $y_{iF}^*,\ i \in S_{nr},$ that minimize
$$\sum_{i \in S}q_i^{-1}\frac{(y_{iF}^*-y_i^*)^2}{y_i^*},$$
subject to (\ref{cal_con}). Straightforward algebra leads to
$$y_{iF}^*=y_i^*\left\{1+q_iw_i\frac{(\widehat{t}_{MR}^*-\widehat{t}_{MR})}{\sum_{i \in S_{nr}}w_i^2q_iy_i^*}\right\}, \quad i \in S_{nr}.$$
The final imputed values $y_{iF}^*$ are readily obtained by using any standard calibration software; e.g., the SAS macro CALMAR2 (Sautory, 2003) and the $R$ package Icarus (Rebecq, 2016).\\

\section{Simulation study}
\begin{figure}[H]
	\centering
	\subfloat[\scriptsize{Normal $\BETA=(10,10,10)$}]{\includegraphics[width=0.3\textwidth]{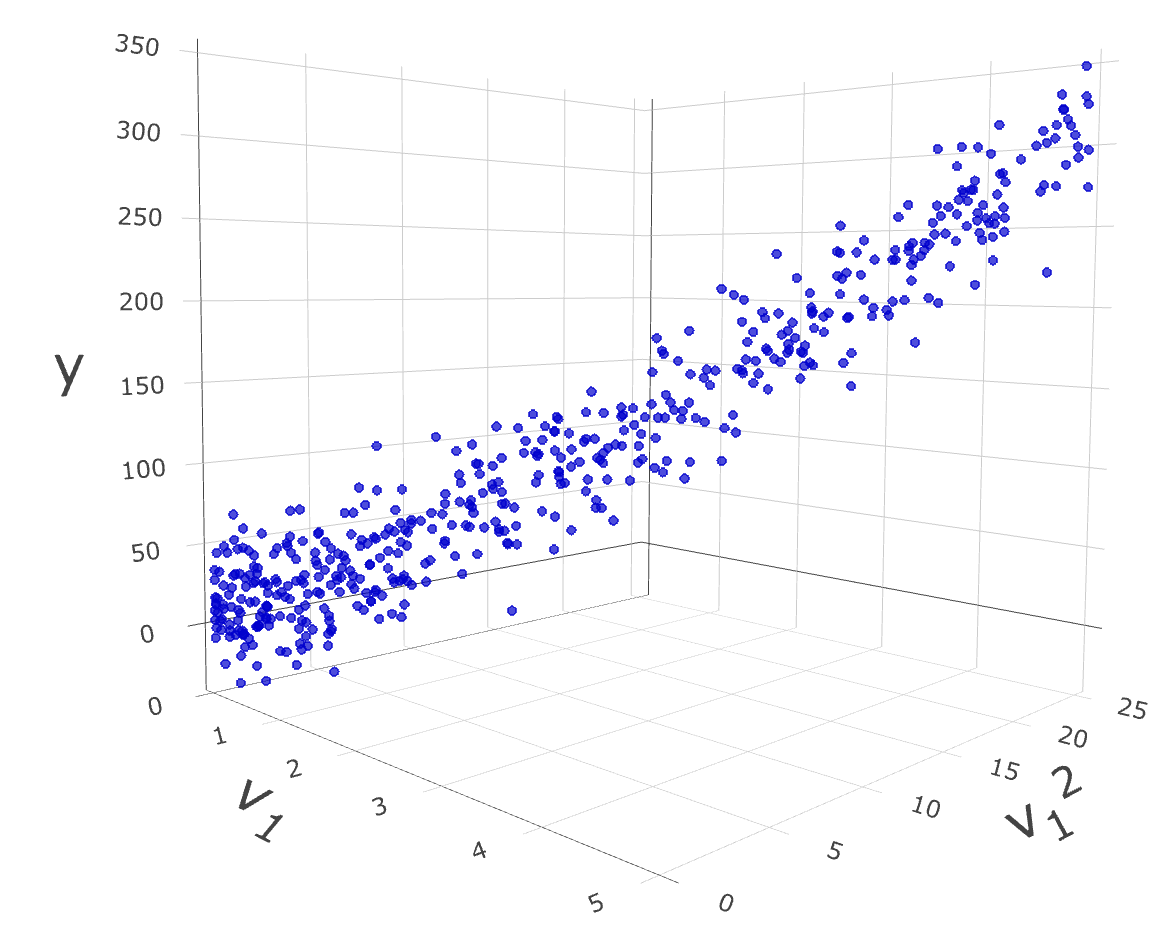}}\\
	\subfloat[\scriptsize{Gamma $\BETA=(1,0.05,0.05)$}]{\includegraphics[width=0.3\textwidth]{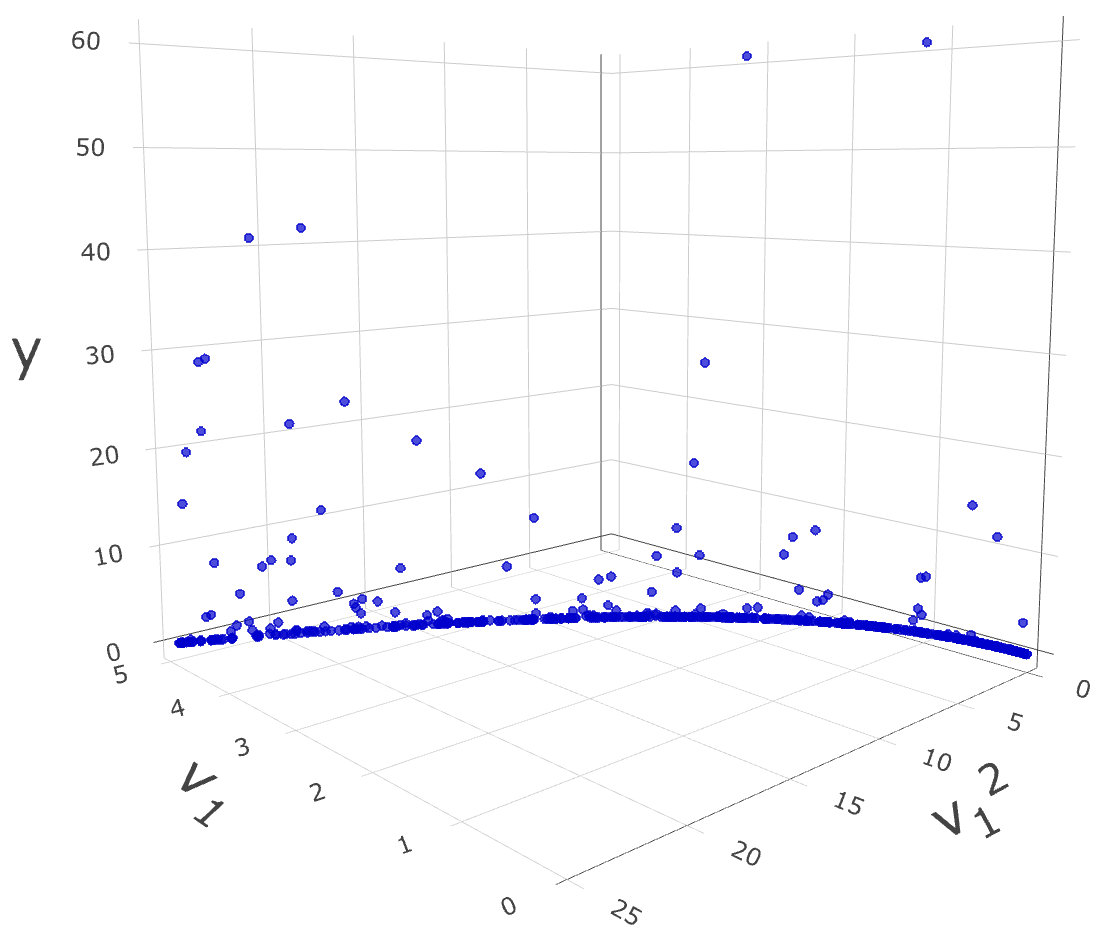}}
	\subfloat[\scriptsize{Gamma $\BETA=(1, 0.2, 0.2)$}]{\includegraphics[width=0.3\textwidth]{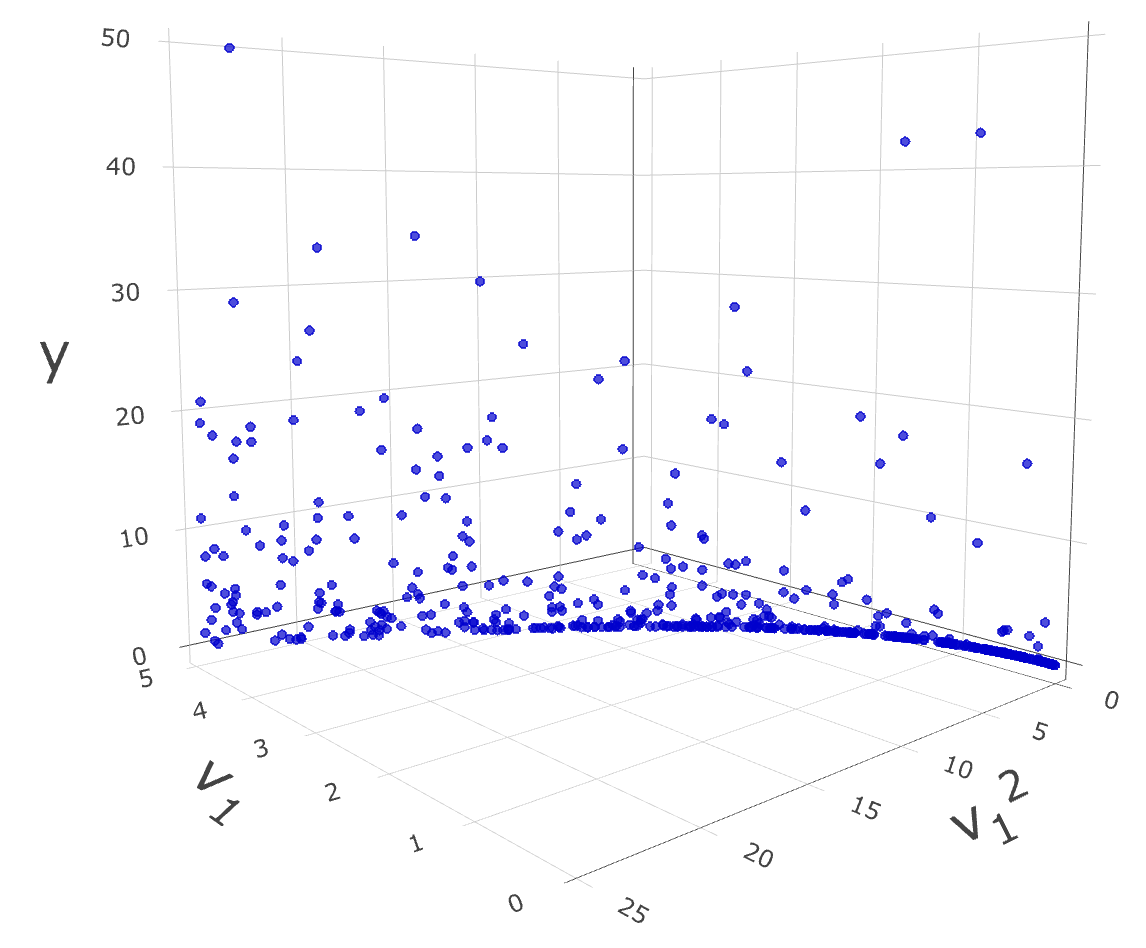}}
	\subfloat[\scriptsize{Gamma $\BETA=(1,1, 0.4)$}]{\includegraphics[width=0.3\textwidth]{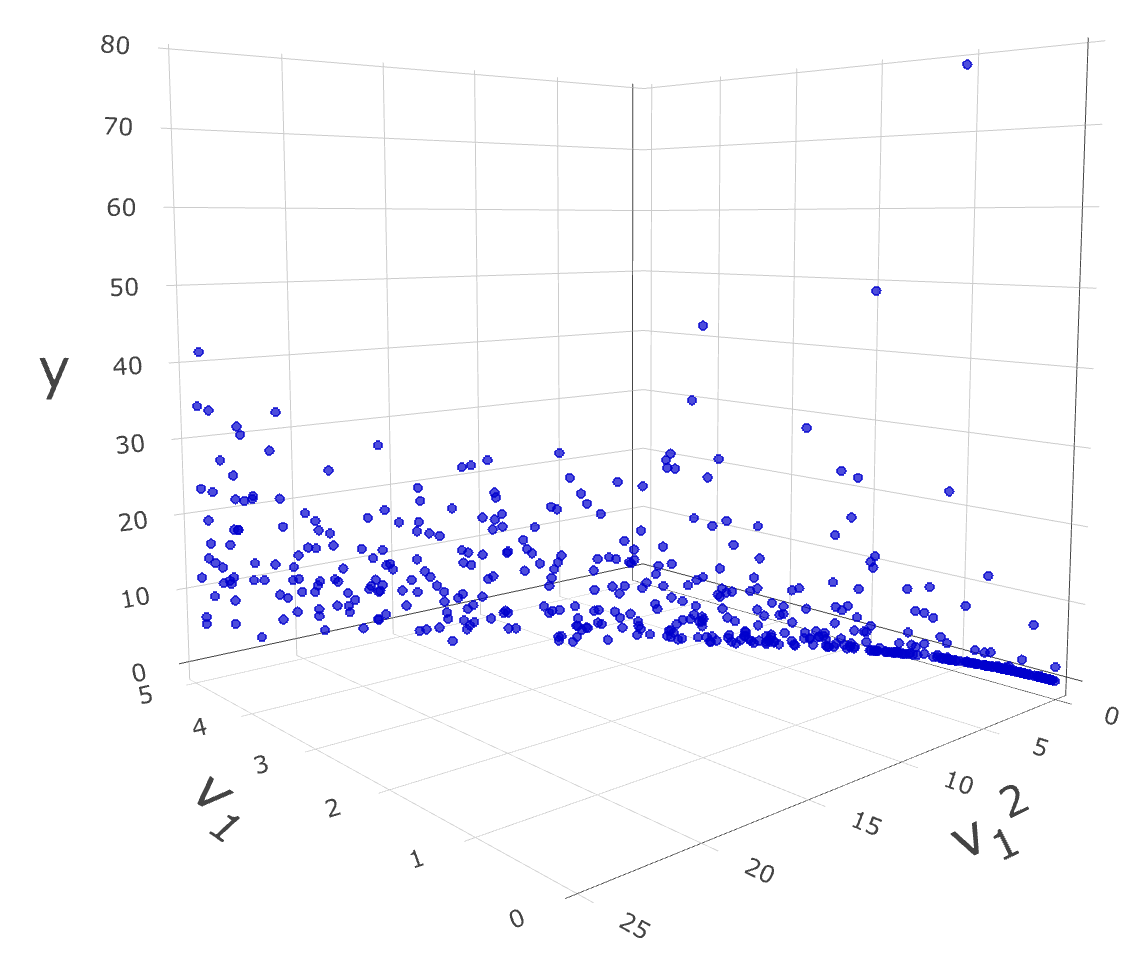}}\\
	\subfloat[\scriptsize{Lognormal $\BETA=(1,0.2, 0.1)$}]{\includegraphics[width=0.3\textwidth]{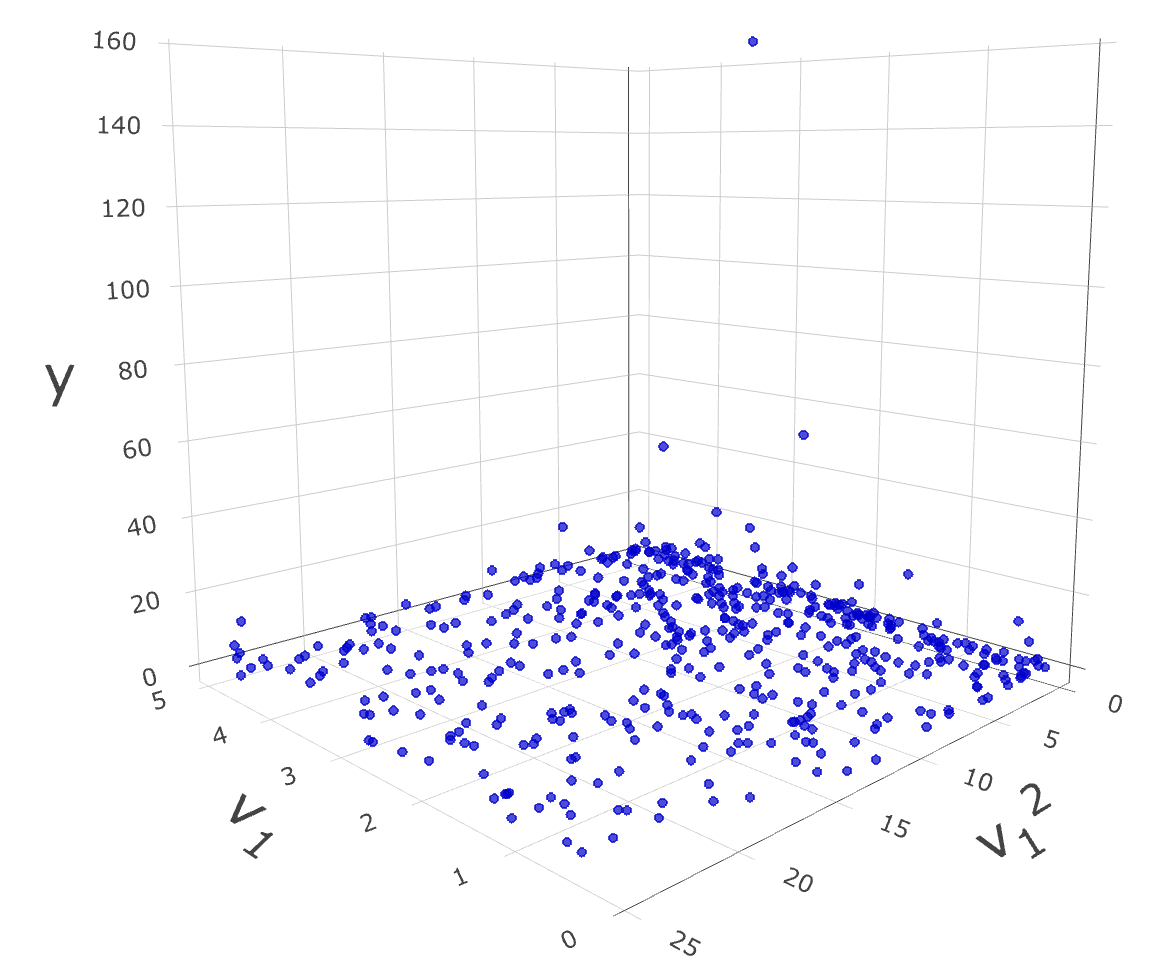}}
	\subfloat[\scriptsize{Lognormal $\BETA=(1,0.3, 0.2)$}]{\includegraphics[width=0.3\textwidth]{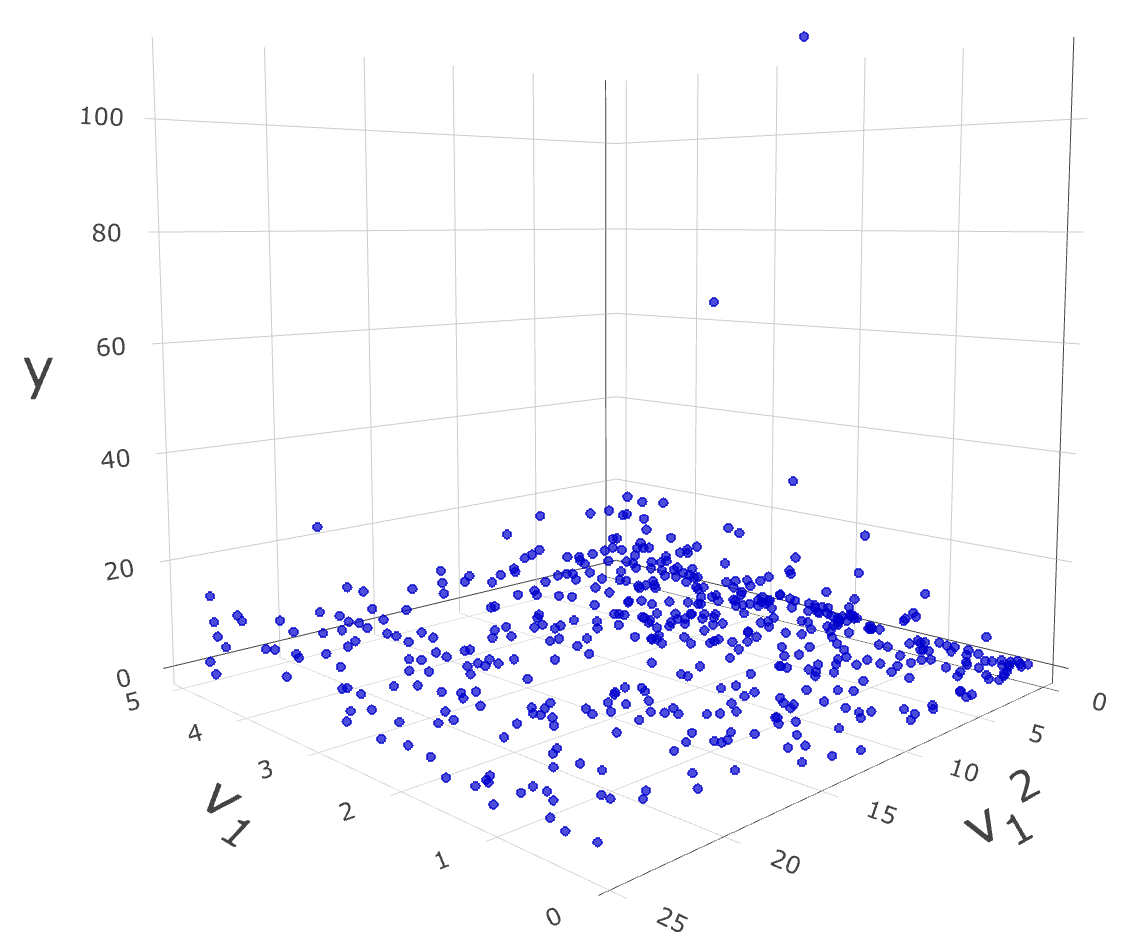}}
	\subfloat[\scriptsize{Lognormal $\BETA=(1,2.3, 0.2)$}]{\includegraphics[width=0.3\textwidth]{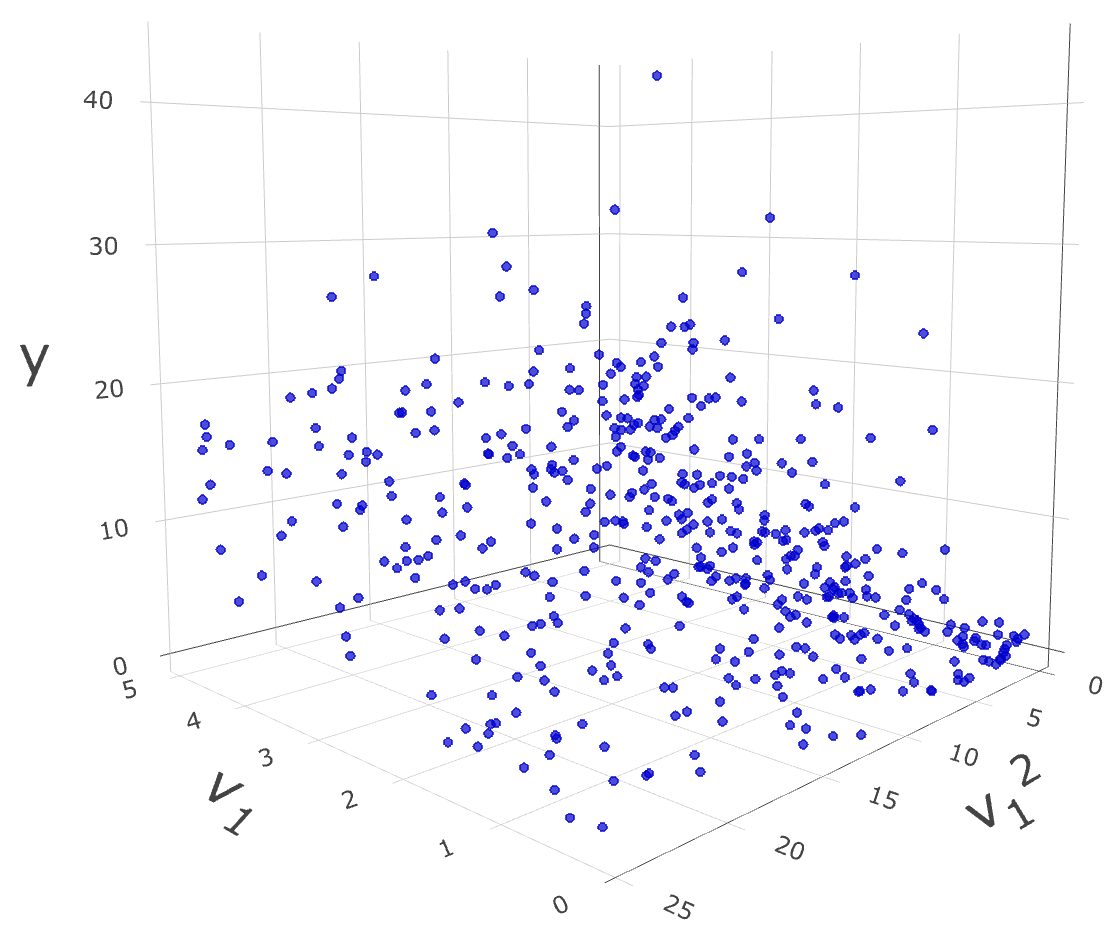}}\\
	\subfloat[\scriptsize{Pareto $\BETA=(1,0.1, 0.1)$}]{\includegraphics[width=0.3\textwidth]{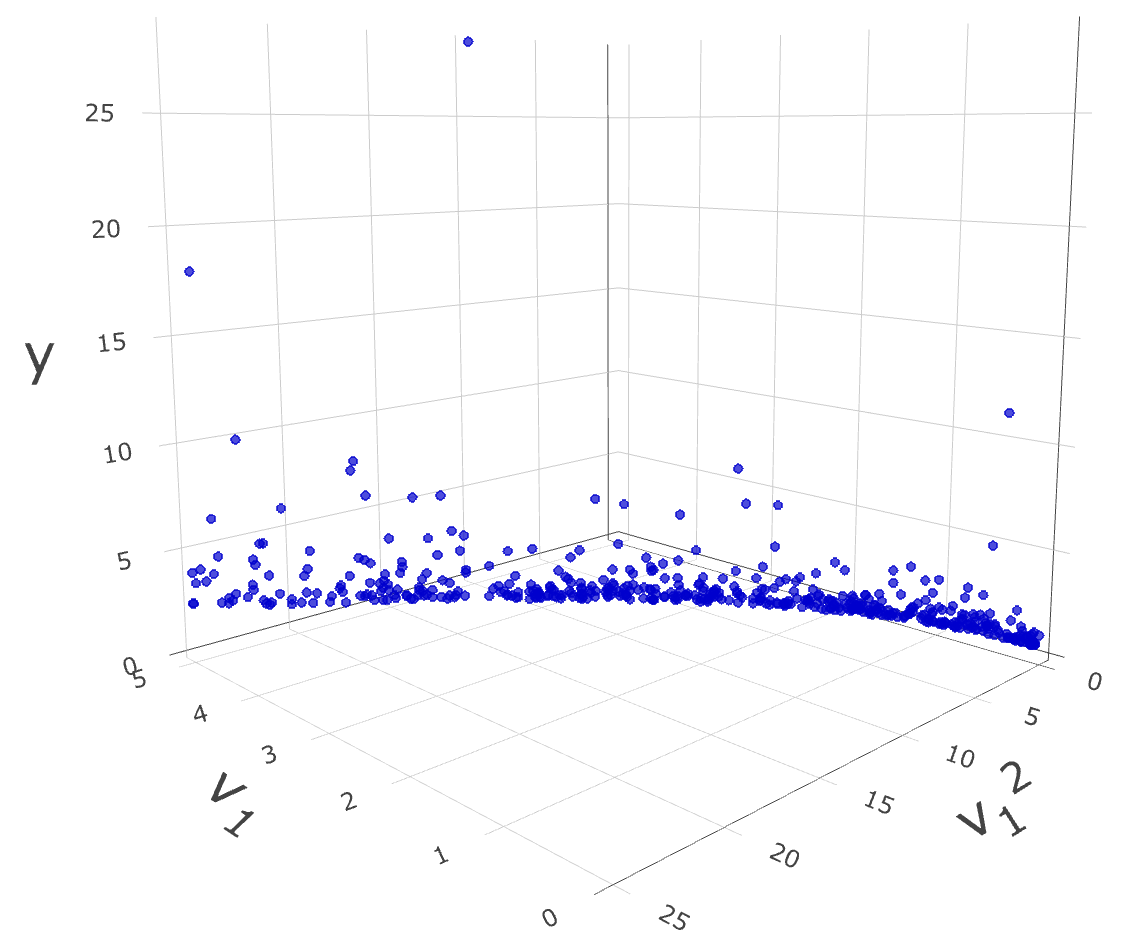}}
	\subfloat[\scriptsize{Pareto $\BETA=(1,0.2, 0.2)$}]{\includegraphics[width=0.3\textwidth]{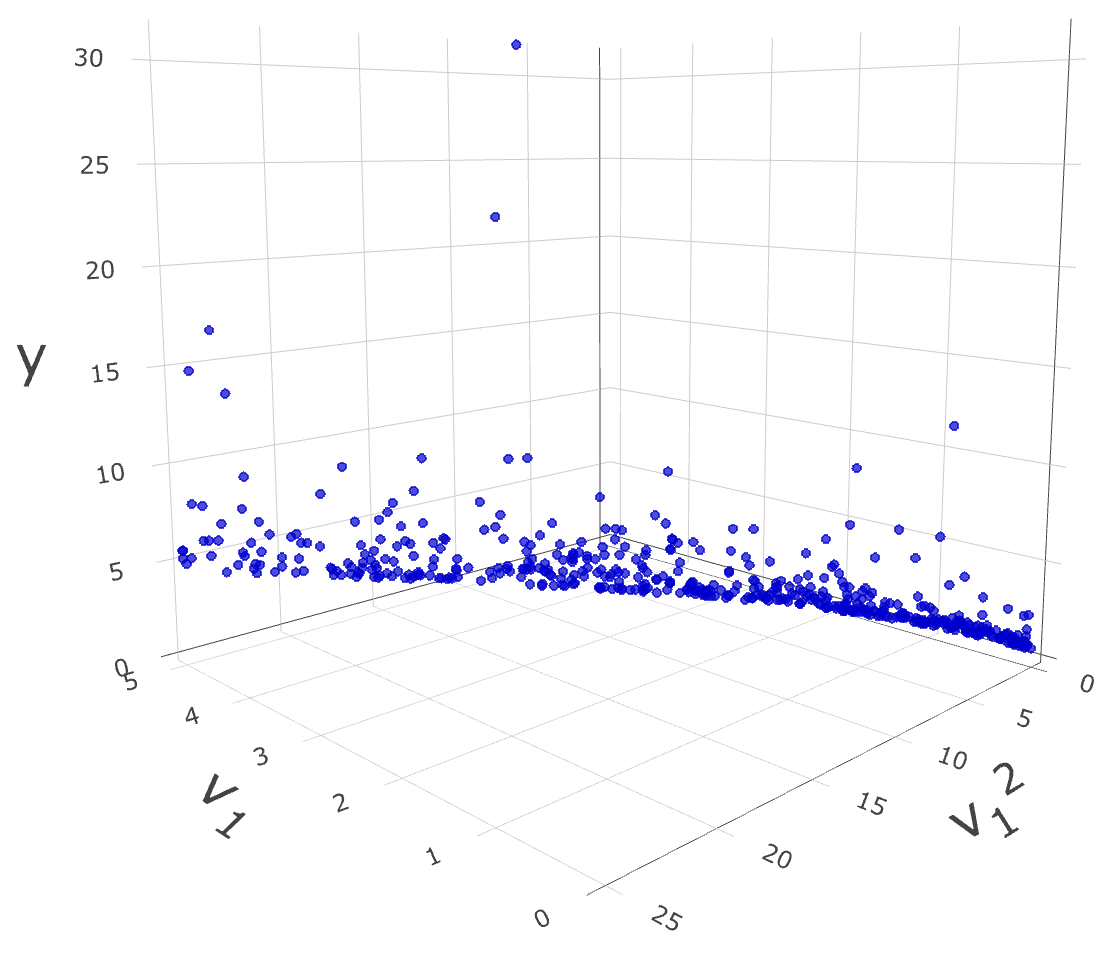}}
	\subfloat[\scriptsize{Pareto  $\BETA=(1,1.5, 0.5)$}]{\includegraphics[width=0.3\textwidth]{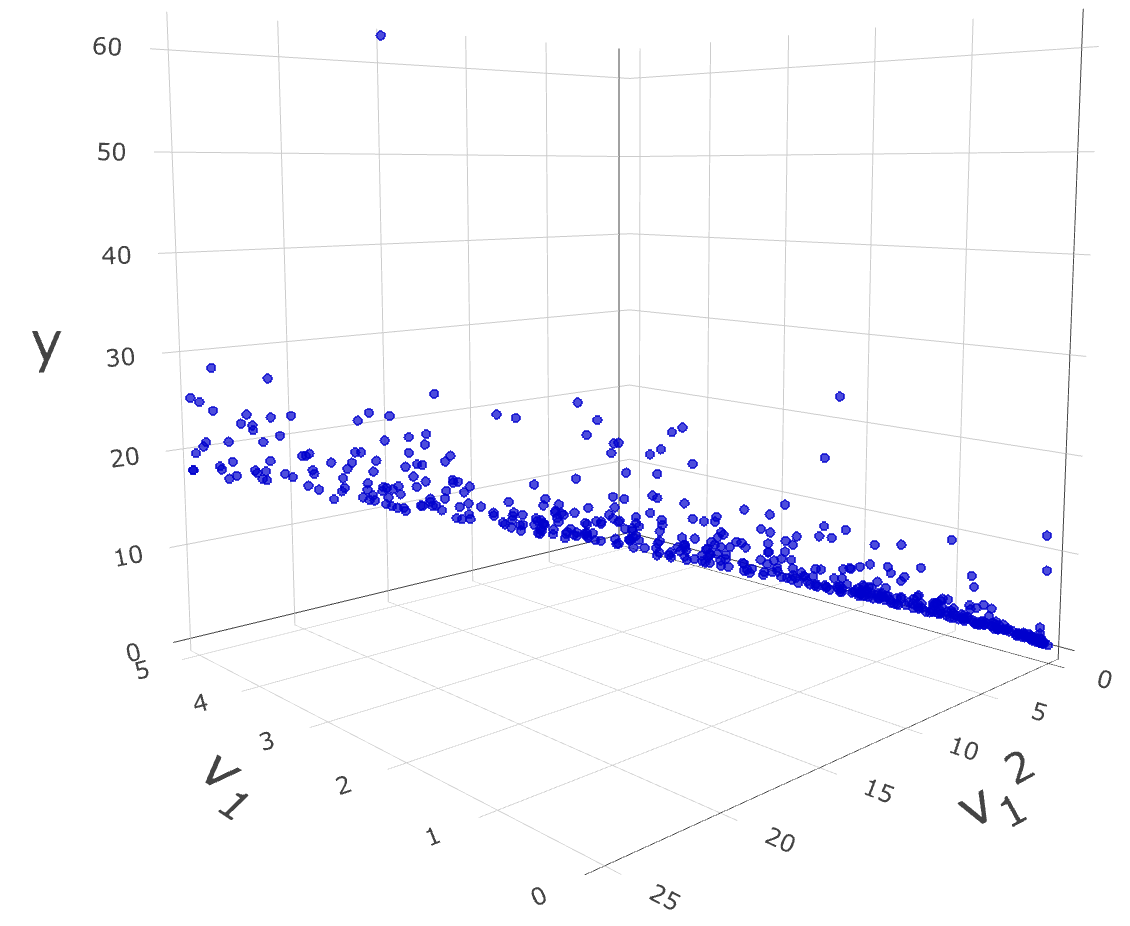}}
	\caption{Four distributions of $y\mid v_{1i}, v_{1i}^2$ }
	\label{fig:10Pop}
\end{figure}

We conducted a simulation study to assess the performance of the proposed method in terms of bias and efficiency. For each scenario, we repeated $R=10, 000$ iterations of the following process:
\begin{itemize}
	\item [(i)] A finite population of size $N=5, 000$ was generated. The population consisted of a survey variable $y$ and a set of predictors. We first generated the auxiliary variable $v_1$ according to a uniform distribution,  $v_1 \sim U(0,5)$.  Given the $v_1$-values, we generated the survey variable $y$ according to four distributions $\mathcal{D}$: normal, Gamma, lognormal and Pareto. More specifically, we used:
	$$y_i \mid v_{1i} \sim \mathcal{D}(\mu_i; \sigma^2_i),$$
	$$\mu_i\equiv \E(y_i \mid v_{1i}, v_{1i}^2)=\beta_0 + \beta_1v_{1i} + \beta_2v_{1i}^2 \quad \mbox{and} \quad \sigma^2_i\equiv \V(y_i \mid v_{1i}, v_{1i}^2) =\sigma^2.$$ 
	Several configurations of the vector $\boldsymbol{\beta}=(\beta_0, \beta_1, \beta_2)$ were used; see the tables of results below. The value of $\sigma^2$ was set to 500, 50, 30 and 20, for the normal, the gamma, the lognormal and the Pareto distributions, respectively. The parameters were set so that the first two moments of the distribution  $\mathcal{D}(\mu_i; \sigma^2_i)$ were the same for the four distributions. Figure 	\ref{fig:10Pop} shows the relationship between $y$ and $\mathbf{v}_i=(v_{1i}, v_{1i}^2)^{\top}$ in each scenario.
	\item [(ii)] From the finite population generated in Step (i), a sample, of size $n=50; 100,$ was selected according to simple random sampling without replacement.

	\item [(iii)] In each sample, the response indicators $r_i,\ i=1, \ldots, n,$ were independently generated according to a Bernoulli distribution with probability 
	$$p_i=\frac{\exp(1.5-1.5v_{1i}+0.4v_{1i}^2)}{1+\exp(1.5-1.5v_{1i}+0.4v_{1i}^2)}.$$ This led to a response rate approximately equal to 70\%.
	\item [(iv)] The missing values in each sample were imputed by three types of imputation procedures:  (a) an imputation based on a single imputation model; (b)  a doubly robust imputation based on a single imputation model and a single nonresponse model; and (c) an imputation based on two imputation models.
	\item [(v)] In each completed data set, the estimators $\widehat{t}_{MR}$ and $\widehat{t}_{MR}^*,$ given respectively by (\ref{Tot_Est_MR}) and (\ref{Tot_Est_MR_Rob}), were computed. 
\end{itemize}
As a measure of bias of an estimator, we computed the Monte Carlo percent relative bias given by
\begin{eqnarray*}
	\E_{MC}(\widehat {t})= \frac{1}{K}\sum_{k=1}^{K}\frac{\widehat{t}_{(k)}-t_y}{t_y}\times 100,	
\end{eqnarray*}
where $\widehat{t}$ is a generic notation used to denote an estimator of $t_y$ and   $\widehat{t}_{(k)}$  is the estimator $\widehat{t}$ at the $k$th iteration. As a measure of efficiency, we computed the percent relative efficiency, using $\widehat{t}_{MR}$ as the reference:
$$\mbox{RE}=100 \times \frac{\mbox{MSE}_{MC}(\widehat{t}_{MR}^*)}{\mbox{MSE}_{MC}(\widehat{t}_{MR})},$$
where
$$\mbox{MSE}_{MC}(\widehat{t})=\frac{1}{K}\sum_{k=1}^{K}(\widehat{t}_{(k)}-t_y)^2.$$

\subsection{Imputation based on a single imputation model}
In this section, the imputed values were constructed using a single imputation model.  In each scenario, we fitted the model $m(\mathbf{v}_i, \BETA)=\mathbf{v}_i^\top\BETA$ with $\mathbf{v}_i=(1, v_i, v_i^2)^{\top}$. That is, the first moment of the imputation was correctly specified.  Table 1 shows the Monte Carlo percent relative bias and relative efficiency of $\widehat{t}_{MR}$ and $\widehat{t}_{MR}^*$ for four distributions.\\

For the normal distribution, the estimator $\widehat{t}_{MR}^*$ showed negligible bias and was slightly less efficient than the estimator  $\widehat{t}_{MR}$ with values of RE equal to 103 for $n=50$ and equal to 101 for $n=100$. For the Gamma distribution, the estimator $\widehat{t}_{MR}^*$  was biased with values of absolute RB ranging from 2.6\% to 22\%. In terms of RE, the estimator $\widehat{t}_{MR}^*$ was more efficient than $\widehat{t}_{MR}$ for all the configurations of the vector $\boldsymbol{\beta}.$ For  $\boldsymbol{\beta}=(1, 0.05, 0.05),$ the estimator $\widehat{t}_{MR}^*$ was much more efficient than $\widehat{t}_{MR}$ with value of RE equal to 67 for $n=50$ and equal to 76 for $n=100$. For the lognormal distribution, the values of absolute RB varied from 1.3\% to 9.7\%. Again, the estimator $\widehat{t}_{MR}^*$ was more efficient than $\widehat{t}_{MR}$ for all the configurations of the vector $\boldsymbol{\beta},$ with values of RE ranging from 67 to 94. Finally, for the Pareto distribution, the estimator $\widehat{t}_{MR}^*$  was moderately biased with values of absolute RB ranging from 1.2\% to 4.7\%. For some configurations of the vector $\boldsymbol{\beta},$ the proposed estimator  $\widehat{t}_{MR}^*$ was considerably more efficient than its counterpart $\widehat{t}_{MR}$; for $\boldsymbol{\beta}=(1, 0.1, 0.1),$ the value of RE was equal to 57. Finally,  the value of RE  with $n=100$ was never less than the value of RE for $n=50.$  

\begin{table}[H]
	\centering
	\begin{tabular}{l cc cc c c}
		\toprule
		Distribution & $\BETA$ & $n$ & $BR_{MC}(\widehat{t}_{MR})$ & $BR_{MC}(\widehat{t}_{MR}^*)$ & $RE$ & \\
		\midrule
		\multirow{2}{*}{Normal} & \multirow{2}{*}{$(10,10,10)$} & $50$ &  $ 0.2$ & $ -0.1$ & $ 103$ \\
		&& $100$ & $ 0.1$ & $ -0.1$ & $ 101$  \\
		\midrule
		\multirow{2}{*}{Gamma} & \multirow{2}{*}{$(1,0.05,0.05)$} & $50$ &  $ -0.3$ & $ -22.9$ & $ 67$ \\
		&& $100$ & $ -0.4$ & $ -17.4$ & $ 76$ \\
		\multirow{2}{*}{Gamma} & \multirow{2}{*}{$(1,0.2,0.2)$} &  $50$ &  $ 0.2$ & $ -10.1$ & $ 82$  \\
		&& $100$ & $ 0.3$ & $ -7.1$ & $ 86$ \\
		\multirow{2}{*}{Gamma} & \multirow{2}{*}{$(1,1,0.4)$} &   $50$ &  $ -0.0$ & $ -3.7$ & $ 94$\\
		&& $100$ & $ -0.1$ & $ -2.6$ & $ 96$ \\
			\midrule
		\multirow{2}{*}{Lognormal} & \multirow{2}{*}{$(1,0.2,0.1)$}  & $50$ &  $ -0.3$ & $ -9.7$ & $ 67$ \\
		&& $100$ & $ -0.2$ & $ -7.2$ & $ 72$ \\
		\multirow{2}{*}{Lognormal} & \multirow{2}{*}{$(1,0.3,0.2)$}  & $50$ &  $ 0.2$ & $ -6.2$ & $ 75$  \\
		&& $100$ & $ 0.1$ & $ -4.5$ & $ 80$ \\
		\multirow{2}{*}{Lognormal} & \multirow{2}{*}{$(1,2.3,0.2)$}  & $50$ &  $ 0.1$ & $ -1.8$ & $ 94$\\
		&& $100$ & $ 0.1$ & $ -1.3$ & $ 94$ \\
		\midrule
		\multirow{2}{*}{Pareto} & \multirow{2}{*}{$(1,0.1,0.1)$} & $50$ &  $ -0.1$ & $ -4.7$ & $ 57$ \\
		&& $100$ & $ 0.2$ & $ -3.4$ & $ 57$ \\
		\multirow{2}{*}{Pareto} & \multirow{2}{*}{$(1,0.2,0.2)$} & $50$ &  $ 0.0$ & $ -4.0$ & $59 $  \\
		&& $100$ & $ 0.0$ & $ -2.9$ & $ 70$ \\
		\multirow{2}{*}{Pareto} & \multirow{2}{*}{$(1,1.5,0.5)$} & $50$ &  $ -0.3$ & $ -2.0$ & $ 92$\\
		&& $100$ & $ 0.0$ & $ -1.2$ & $ 92$ \\
		\bottomrule
	\end{tabular}
	\caption{Monte Carlo percent relative bias and relative efficiency of  $\widehat{t}_{MR}$ and $\widehat{t}_{MR}^*$ for four distributions}
\end{table}

\subsection{Imputation based on a single imputation model and a single nonresponse model}
\label{RES_imp1m1p}

In this section, the finite populations and the nonresponse indicators were generated using the same models as in Section 6.1. We considered the case of doubly robust imputation procedures for which the imputer specifies an imputation model and a nonresponse model.  In the three scenarios described below, we fitted a nonresponse model and an imputation model of the form 
$$p(\mathbf{v}_i, \ALPHA)=\frac{\exp(\mathbf{v}_i^\top\ALPHA)}{1+\exp(\mathbf{v}_i^\top\ALPHA)} \quad \mbox{and} \quad m(\mathbf{v}_i, \BETA)=\mathbf{v}_i^\top\BETA.$$
We considered three scenarios:
\begin{itemize}
	\item[(i)] Both models were correctly specified, denoted by $m$ $\blacksquare$ and  $p$ $\blacksquare$;
	\item[(ii)] The imputation model was correctly specified but the nonresponse model was misspecified, denoted by $m$ $\blacksquare$ and $p$ $\square$;
	\item[(iii)] The nonresponse model was correctly specified but the imputation model was misspecified, denoted by $m$ $\square$ and $p$ $\blacksquare$. 
\end{itemize}

Correctly specified models were based on the set of predictors 1, $v_1$ and $v_1^2$. Misspecified models were based on the set of predictors  2, $v_1$ and $v_2,$ where $v_2$ was generated from a $\mathcal{U}(0,4)$ and was unrelated to both the survey variable $y$ and the response indicators $r$. This is summarized in Table \ref{tab:modeles_specification1m1p}.

\begin{table}[H]
	\center
	\begin{tabular}{c c c c}
		\toprule
		& $v_1$ & $v_1^2$ & $v_2$ \\
		\hline
		$m$ $\blacksquare$ & \multirow{2}{*}{\checkmark} & \multirow{2}{*}{\checkmark} & \multirow{2}{*}{X} \\
		$p$ $\blacksquare$ & \\
		$m$ $\square$ & \multirow{2}{*}{\checkmark} & \multirow{2}{*}{X} & \multirow{2}{*}{\checkmark} \\
		$p$ $\square$ & \\
		\bottomrule
	\end{tabular}
	\caption{Summary of the three scenarios}
	\label{tab:modeles_specification1m1p}
\end{table}

\begin{table}[H]
	\centering
	\begin{tabular}{c cc c c c}
		\toprule
		$\BETA$ & Scenario & $n$ & $BR_{MC}(\widehat{t}_{MR})$ & $BR_{MC}(\widehat{t}_{MR}^R)$ & $RE$  \\
		\midrule
		\multirow{6}{*}{$(10, 10, 10)$} &  \multirow{2}{*}{ $m$ $\blacksquare$ , $p$ $\blacksquare$ } & 50 &  $ 0.1$ & $ -0.3$ & $ 103$ \\
		&& 100 &  $ -0.1$ & $ -0.3$ & $ 101$ \\
		& \multirow{2}{*}{$m$ $\blacksquare$ , $p$ $\square$} & 50 &  $ -0.1$ & $ -0.4$ & $ 103$ \\
		&& 100 &  $ -0.0$ & $ -0.2$ & $ 102$ \\
		& \multirow{2}{*}{$m$ $\square$ , $p$ $\blacksquare$}  & 50 &  $ 2.2$ & $ 1.9$ & $ 101$ \\
		&& 100 &  $ 2.2$ & $ 2.0$ & $ 100$ \\
		\bottomrule
	\end{tabular}
	\caption{Monte Carlo percent relative bias and relative efficiency of  $\widehat{t}_{MR}$ and $\widehat{t}_{MR}^*$ for the normal distribution}
\end{table}

\begin{table}[H]
	\centering
	\begin{tabular}{c c cc c c}
		\toprule
		$\BETA$ & Scenario & $n$ & $BR_{MC}(\widehat{t}_{MR})$ & $BR_{MC}(\widehat{t}_{MR}^R)$ & $RE$  \\
		\midrule
		\multirow{6}{*}{$(1,0.05,0.05)$}  & \multirow{2}{*}{$m$ $\blacksquare$ , $p$ $\blacksquare$}  &  50 &  $ 0.6$ & $ -21.7$ & $ 68$ \\
		&& 100 &  $ -0.2$ & $ -17.4$ & $ 76$ \\
		&\multirow{2}{*}{$m$ $\blacksquare$ , $p$ $\square$} & 50 &  $ 0.2$ & $ -21.8$ & $ 69$ \\
		&& 100 &  $ -0.3$ & $ -17.0$ & $ 78$ \\
		&\multirow{2}{*}{$m$ $\square$ , $p$ $\blacksquare$} & 50 &  $ 0.1$ & $ -21.4$ & $ 72$ \\
		&& 100 &  $ 1.2$ & $ -15.5$ & $ 78$ \\
		\midrule
		\multirow{6}{*}{$(1,0.2,0.2)$} & \multirow{2}{*}{$m$ $\blacksquare$ , $p$ $\blacksquare$}  & 50 &  $ -0.1$ & $ -10.3$ & $ 83$ \\
		&& 100 &  $ 0.1$ & $ -7.2$ & $ 86$ \\
		&\multirow{2}{*}{$m$ $\blacksquare$ , $p$ $\square$}  & 50 &  $ 0.4$ & $ -9.6$ & $ 83$ \\
		&& 100 &  $ 0.6$ & $ -6.5$ & $ 86$ \\
		&\multirow{2}{*}{$m$ $\square$ , $p$ $\blacksquare$}  & 50 &  $ 2.1$ & $ -7.8$ & $ 82$ \\
		&& 100 &  $ 2.2$ & $ -5.0$ & $ 83$ \\
		\midrule
		\multirow{6}{*}{$(1,1,0.4)$} &\multirow{2}{*}{$m$ $\blacksquare$ , $p$ $\blacksquare$}  & 50 &  $ -0.2$ & $ -3.7$ & $ 94$ \\
		&& 100 &  $ -0.1$ & $ -2.6$ & $ 95$ \\
		&\multirow{2}{*}{$m$ $\blacksquare$ , $p$ $\square$}  & 50 &  $ 0.2$ & $ -3.4$ & $ 95$ \\
		&& 100 &  $ 0.0$ & $ -2.5$ & $ 95$ \\
		&\multirow{2}{*}{$m$ $\square$ , $p$ $\blacksquare$}  & 50 &  $ 1.1$ & $ -2.3$ & $ 92$ \\
		&& 100 &  $ 1.7$ & $ -0.7$ & $ 90$ \\
		\bottomrule
	\end{tabular}
	\caption{Monte Carlo percent relative bias and relative efficiency of  $\widehat{t}_{MR}$ and $\widehat{t}_{MR}^*$ for the Gamma distribution}
\end{table}

\begin{table}[H]
	\centering
	\begin{tabular}{c ccc c c}
		\toprule
		$\BETA$ & Scenario & $n$ & $BR_{MC}(\widehat{t}_{MR})$ & $BR_{MC}(\widehat{t}_{MR}^R)$ & $RE$  \\
		\midrule
		\multirow{6}{*}{$(1,0.2,0.1)$ }   & \multirow{2}{*}{$m$ $\blacksquare$ , $p$ $\blacksquare$}  &  50 &  $ 0.6$ & $ -8.7$ & $ 64$ \\
		&& 100 &  $ 0.1$ & $ -7.0$ & $ 68$ \\
		&\multirow{2}{*}{$m$ $\blacksquare$ , $p$ $\square$}   & 50 &  $ 0.2$ & $ -9.1$ & $ 68$ \\
		&& 100 &  $ -0.0$ & $ -6.9$ & $ 71$ \\
		&\multirow{2}{*}{$m$ $\square$ , $p$ $\blacksquare$}   & 50 &  $ 0.5$ & $ -8.3$ & $ 70$ \\
		&& 100 &  $ 1.1$ & $ -5.8$ & $ 67$ \\
		\midrule
		\multirow{6}{*}{$(1,0.3,0.2)$ }   & \multirow{2}{*}{$m$ $\blacksquare$ , $p$ $\blacksquare$}  & 50 & $ 0.2$ & $ -6.1$ & $ 71$  \\
		&& 100 &   $ -0.1$ & $ -4.6$ & $ 79$\\
		&\multirow{2}{*}{$m$ $\blacksquare$ , $p$ $\square$}   & 50 &  $ -0.4$ & $ -6.6$ & $ 79$ \\
		&& 100 &  $ 0.0$ & $ -4.4$ & $ 78$ \\
		&\multirow{2}{*}{$m$ $\square$ , $p$ $\blacksquare$}   & 50 &  $ 1.4$ & $ -4.6$ & $ 68$ \\
		&& 100 &  $ 1.7$ & $ -2.8$ & $ 76$ \\
		\midrule
		\multirow{6}{*}{$(1,2.3,0.2)$ }   & \multirow{2}{*}{$m$ $\blacksquare$ , $p$ $\blacksquare$}  & 50 &  $ 0.1$ & $ -1.7$ & $ 92$ \\
		&& 100 &  $ 0.1$ & $ -1.2$ & $ 94$ \\
		&\multirow{2}{*}{$m$ $\blacksquare$ , $p$ $\square$}   & 50 &  $ 0.1$ & $ -1.8$ & $ 93$ \\
		&& 100 &  $ -0.1$ & $ -1.4$ & $ 95$ \\
		&\multirow{2}{*}{$m$ $\square$ , $p$ $\blacksquare$}   & 50 &  $ 0.4$ & $ -1.4$ & $ 92$ \\
		&& 100 &  $ 0.6$ & $ -0.7$ & $ 93$\\
		\bottomrule
	\end{tabular}
	\caption{Monte Carlo percent relative bias and relative efficiency of  $\widehat{t}_{MR}$ and $\widehat{t}_{MR}^*$ for the lognormal distribution}
\end{table}

\begin{table}[H]
	\centering
	\begin{tabular}{c ccc c c}
		\toprule
		$\BETA$ & Scenario & $n$ & $BR_{MC}(\widehat{t}_{MR})$ & $BR_{MC}(\widehat{t}_{MR}^R)$ & $RE$  \\
		\midrule
		\multirow{6}{*}{$(1,0.1,0.1)$}  & \multirow{2}{*}{$m$ $\blacksquare$ , $p$ $\blacksquare$}  &  50 &  $ -0.2$ & $ -4.7$ & $ 56$ \\
		&& 100 &  $ -0.1$ & $ -3.6$ & $ 63$ \\
		&\multirow{2}{*}{$m$ $\blacksquare$ , $p$ $\square$}   & 50 &  $ 0.3$ & $ -4.3$ & $ 56$ \\
		&& 100 &  $ 0.1$ & $ -3.4$ & $ 59$ \\
		&\multirow{2}{*}{$m$ $\square$ , $p$ $\blacksquare$}   & 50 &  $ 1.0$ & $ -3.4$ & $ 53$ \\
		&& 100 &  $ 1.5$ & $ -2.1$ & $ 53$ \\
		\midrule
		\multirow{6}{*}{$(1,0.2,0.2)$}  & \multirow{2}{*}{$m$ $\blacksquare$ , $p$ $\blacksquare$}  & 50 &  $ 0.0$ & $ -3.9$ & $ 68$ \\
		&& 100 &  $ 0.1$ & $ -2.9$ & $ 67$ \\
		&\multirow{2}{*}{$m$ $\blacksquare$ , $p$ $\square$ }  & 50 &  $ 0.3$ & $ -3.6$ & $ 66$ \\
		&& 100 &  $ -0.2$ & $ -3.1$ & $ 77$ \\
		&\multirow{2}{*}{$m$ $\square$ , $p$ $\blacksquare$}   & 50 &  $ 1.8$ & $ -1.9$ & $ 67$ \\
		&& 100 &  $ 1.7$ & $ -1.2$ & $ 67$ \\
		\midrule
		\multirow{6}{*}{$(1,1.5,0.5)$} & \multirow{2}{*}{$m$ $\blacksquare$ , $p$ $\blacksquare$}  & 50 &  $ 0.0$ & $ -1.7$ & $ 91$ \\
		&& 100 &  $ -0.0$ & $ -1.2$ & $ 91$ \\
		&\multirow{2}{*}{$m$ $\blacksquare$ , $p$ $\square$}   & 50 &  $ 0.1$ & $ -1.6$ & $ 92$ \\
		&& 100 &  $ -0.0$ & $ -1.2$ & $ 93$ \\
		&\multirow{2}{*}{$m$ $\square$ , $p$ $\blacksquare$ }  & 50 &  $ 1.3$ & $ -0.3$ & $ 88$ \\
		&& 100 &  $ 1.5$ & $ 0.3$ & $ 88$ \\
		\bottomrule
	\end{tabular}
	\caption{Monte Carlo percent relative bias and relative efficiency of  $\widehat{t}_{MR}$ and $\widehat{t}_{MR}^*$ for the Pareto distribution}
\end{table}

The results are shown in Tables 3-6. As expected, the estimator $\widehat{t}_{MR}$ showed a small bias in all the scenarios. This can be explained by the fact that it is doubly robust in the sense that it remains consistent for the true total $t_y$ if either model is correctly specified. The results in Tables 3-6 were similar to those obtained in Section 6.1. The estimator $\widehat{t}_{MR}^*$ was biased but more efficient than $\widehat{t}_{MR}$ in all the scenarios. Again, the gains in efficiency were especially noteworthy for the Pareto distribution with values of RE ranging from 53 to 92; see Table 6.  

\subsection{Imputation based on two imputation models}
\label{RES_Imp2m0p}
Again, the finite populations and the nonresponse indicators were generated using the same models as in Section 6.1.
In this section, the imputed values were based on two imputation models:
 $$m^{(1)}(\mathbf{v}_i^{(1)}, \BETA^{(1)})=\mathbf{v}_i^{(1)\top}\BETA^{(1)} \quad \mbox{and} \quad m^{(2)}(\mathbf{v}_i^{(2)}, \BETA^{(2)})=\mathbf{v}_i^{(2)\top}\BETA^{(2)}.$$ 
The model $m^{(1)}(\mathbf{v}_i^{(1)}, \BETA^{(1)})$ was correctly specified, whereas the model $m^{(2)}(\mathbf{v}_i^{(2)}, \BETA^{(2)})$ was misspecified. Table \ref{tab:modeles_specification2m0p} gives the set of predictors for each model.  
\begin{table}[H]
	\center
	\begin{tabular}{c c c c}
		\toprule
		& $v_1$ & $v_1^2$ & $v_2$ \\
		\hline
		$m^{(1)} \blacksquare$ & \checkmark & \checkmark & X \\
		\addlinespace
		$m^{(2)} \square$ & \checkmark & X & \checkmark \\
		\bottomrule
	\end{tabular}
	\caption{Working models}
	\label{tab:modeles_specification2m0p}
\end{table}

The results are shown in Table 8. Again, the results were very similar to those obtained in Sections 6.1 and 6.2.

\begin{table}[H]
	\centering
	\begin{tabular}{l cc c c c}
		\toprule
		Distribution & $\BETA$ & $n$ & $BR_{MC}(\widehat{t}_{MR})$ & $BR_{MC}(\widehat{t}_{MR}^R)$ & $RE$ \\
		\midrule
		\multirow{2}{*}{Normal} & \multirow{2}{*}{$(10,10,10)$} & 50 &  $0.0$ & $-0.5$ & $103$   \\
		&& 100 &  $ -0.0$ & $ -0.3$ & $ 102$   \\
		\midrule
		\multirow{2}{*}{Gamma} & \multirow{2}{*}{$(1,0.05,0.05)$ } & 50 &  $ 0.7$ & $ -15.9$ & $ 76$   \\
		&& 100 &  $ 0.9$ & $ -11.4$ & $ 81$   \\
		\multirow{2}{*}{Gamma} & \multirow{2}{*}{$(1,0.2,0.2)$} & 50 &  $ 0.1$ & $ -7.4$ & $ 88$   \\
		&& 100 &  $ 0.5$ & $ -4.5$ & $ 90$   \\
		\multirow{2}{*}{Gamma} & \multirow{2}{*}{$(1,1,0.4)$}  & 50 &  $ 0.5$ & $ -2.2$ & $ 95$   \\
		&& 100 &  $ 0.3$ & $ -1.4$ & $ 97$   \\
		\midrule
		\multirow{2}{*}{Lognormal} & \multirow{2}{*}{$(1,0.2,0.1)$} & 50 &  $ 0.4$ & $ -6.9$ & $ 75$   \\
		&& 100 &  $ 0.2$ & $ -5.0$ & $ 78$   \\
		\multirow{2}{*}{Lognormal} & \multirow{2}{*}{$(1,0.3,0.2)$} & 50 &  $ 0.5$ & $ -4.4$ & $ 81$   \\
		&& 100 &  $ 0.3$ & $ -3.1$ & $ 86$   \\
		\multirow{2}{*}{Lognormal} & \multirow{2}{*}{$(1,2.3,0.2)$} & 50 &  $ 0.1$ & $ -1.3$ & $ 97$   \\
		&& 100 &  $ -0.1$ & $ -1.1$ & $ 98$   \\
		\midrule
		\multirow{2}{*}{Pareto} & \multirow{2}{*}{$(1,0.1,0.1)$} & 50 &  $ 0.4$ & $ -3.3$ & $ 63$   \\
		&& 100 &  $ 0.4$ & $ -2.4$ & $ 69$   \\
		\multirow{2}{*}{Pareto} & \multirow{2}{*}{$(1,0.2,0.2)$}& 50 &  $ 0.4$ & $ -2.8$ & $ 73$   \\
		&& 100 &  $ 0.1$ & $ -2.2$ & $ 76$   \\
		\multirow{2}{*}{Pareto} & \multirow{2}{*}{$(1,1.5,0.5)$} & 50 &  $ 0.2$ & $ -1.1$ & $ 93$   \\
		&& 100 &  $ 0.0$ & $ -0.9$ & $ 92$   \\ 
		\bottomrule
	\end{tabular}
	\caption{Monte Carlo percent relative bias and relative efficiency of  $\widehat{t}_{MR}$ and $\widehat{t}_{MR}^*$ for four distributions}
\end{table}

\section{Final remarks}
In this paper, we have proposed an efficient version of the customary multiply robust estimator based on the concept of conditional bias of a unit. The proposed method is general as it can be applied to a wide class of imputation procedures including the customary imputation based on a single imputation model and doubly robust imputation procedures. The results from a simulation study suggest that the proposed method outperforms the customary multiply robust estimator in terms of mean square error when the distribution of $y$ given $\mathbf{v}$ is highly skewed. The gains were especially substantial in the case of the lognormal and the Pareto distributions.\\

It would be of interest to develop an estimator of the mean square error of the proposed estimator $\widehat{t}_{MR}^*$ to assess its efficiency in practice.  A satisfactory solution to this issue is currently lacking, even in the ideal case of 100\% response. Although a bootstrap procedure would seem natural, the extreme order statistics $\widehat{B}_{\min}^{(MR)}$ and $\widehat{B}_{\max}^{(MR)}$ in $\widehat{t}_{MR}^*$ make the application of bootstrap relatively complex. This issue will be considered elsewhere.

\section*{Acknowledgment}
The first author's research was partially supported by the Oklahoma Shared Clinical and Translational Resources (U54GM104938) with an Institutional Development Award (IDeA) from NIGMS. The content is solely the responsibility of the authors and does not necessarily represent the official views of the National Institutes of Health. The second author wishes to acknowledge the support of grants from the Natural Sciences and Engineering
Research Council of Canada. The third author wishes to acknowledge the support of grants from the Canadian Statistical Sciences Institute.

\bigskip
\centerline{ {\sc References} }
\addcontentsline{toc}{section}{References}

\begin{description}
	\item Beaumont, J.-F. (2005). Calibrated imputation in surveys under a quasi-model-assisted approach. \textit{Journal of the Royal Statistical Society: Series B (Statistical Methodology)}, \textbf{67}:445--458.
	
	\item Beaumont, J.-F. and Alavi, A. (2004). Robust generalized regression estimation. \textit{Survey Methodology}, \textbf{30}:195--208.
	
	\item Beaumont, J.-F., Haziza, D., and Ruiz-Gazen, A. (2013). A unified approach to robust estimation in finite population sampling. \textit{Biometrika}, \textbf{100}:555--569. 
	
	\item Breidt, F. J., Opsomer, J. D., et al. (2017). Model-assisted survey estimation with modern prediction techniques. \textit{Statistical Science}, \textbf{32}:190--205.
	
	\item Cao, W., Tsiatis, A. A. and Davidian, M. (2009). Improving efficiency and robustness of the doubly robust estimator for a population mean with incomplete data. \textit{Biometrika}, \textbf{96}:723--734.
	
	\item Chambers, R. L. (1986). Outlier robust finite population estimation. \textit{Journal of the American Statistical Association}, \textbf{81}:1063--1069.
	
	\item Chan, K. C. G. and Yam, S. C. P. (2014). Oracle, multiple robust and multipurpose calibration in a missing response problem. \textit{Statistical Science}, \textbf{29}:380--396. 
	
	\item Chen, S. and Haziza, D. (2017). Multiply robust imputation procedures for the treatment of item nonresponse in surveys. \textit{Biometrika}, \textbf{104}:439--453. 
	
	\item Chen, S. and Haziza, D. (2019). Recent developments in dealing with item non-response in surveys: A critical review. \textit{International Statistical Review}, \textbf{87}:S192--S218.
	
	\item Deville, J.-C. and Särndal, C.-E. (1992). Calibration estimators in survey
	sampling. \textit{Journal of the American Statistical Association} \textbf{87}: 376--382.
	
	\item Dongmo Jiongo, V. (2015). \textit{Inférence robuste à la présence des valeurs aberrantes dans les enquêtes.} PhD thesis, Université de Montréal.
	
	\item Favre-Martinoz, C., Haziza, D. and Beaumont, J.-F.  (2016). Robust inference in two-phase sampling designs with application to unit nonresponse. \textit{Scandinavian Journal of Statistics}, \textbf{43}:1019--1034.
	
	\item Han, P. (2014a). A further study of the multiply robust estimator in missing data analysis. \textit{Journal of Statistical Planning and Inference}, \textbf{148}:101--110.
	
	\item Han, P. (2014b). Multiply robust estimation in regression analysis with missing data. \textit{Journal of the American Statistical Association}, \textbf{109}:1159--1173.
	
	\item Han, P. and Wang, L. (2013). Estimation with missing data: beyond double robustness. \textit{Biometrika}, \textbf{100}:417--430.
	
	
	\item Haziza, D. and Rao, J. N. K. (2006). A nonresponse model approach to inference under imputation for missing survey data. \textit{Survey Methodology}, \textbf{32}:53--64. 
	
	\item Horvitz, D. G. and Thompson, D. J. (1952). A generalization of sampling without replacement from a finite universe. \textit{Journal of the American statistical Association}, \textbf{47}:663--685.
	
	\item Kang, J. D. Y. and Schafer, J. L. (2007). Demystifying double robustness: a comparison of alternative strategies for estimating a population mean from incomplete data. \textit{Statistical Science}, \textbf{22}:523--539
	
	\item Kim, J. K. and Haziza, D. (2014). Doubly robust inference with missing data in survey sampling. \textit{Statistica Sinica}, \textbf{24}:375--394. 
	
	\item Kim, J. K. and Park, H. A. (2006). Imputation using response probability. \textit{Canadian Journal of Statistics}, \textbf{34}:171--182. 
	
	\item Mashreghi, Z., Haziza, D., Léger, C. (2016). A survey of bootstrap methods in finite population sampling. \textit{Statistics Surveys}, \textbf{10}:1--52.
	
	\item Moreno-Rebollo, J.~L., Mu\~{n}oz-Reyez, A.~M., Jiménez-Gamero, M.~D. and Mu\~{n}oz-Pichardo, J. (2002). Influence diagnostics in survey sampling: estimating the conditional bias. {\em Metrika}, \textbf{55}:209--214.
	
	\item  Moreno-Rebollo, J. L., Mu\~{n}oz-Reyez, A. M. and Mu\~{n}oz-Pichardo, J. M. (1999). Miscellanea. Influence diagnostics in survey sampling: conditional bias.
	{\em Biometrika}, \textbf{86}:923--928.
	
	\item  Mu\~{n}oz-Pichardo, J., Mu\~{n}oz-Garcia, J., Moreno-Rebollo, J.~L. and Pi\~{n}o-Mejias, R.  (1995). A new approach to influence analysis in linear models.
	{\em Sankhya, Series A}, \textbf{57}:393--409.
	
	\item Rebecq, A. (2016). “Icarus: an R package for calibration in survey sampling.” R package version 0.2.0.
	
	\item Ren, R. and Chambers, R. (2003). Outlier robust imputation of survey data via reverse calibration.
	
	\item 	Robins, J. M., Rotnitzky, A. and Zhao, L. P. (1994). Estimation of regression coefficient when some regressors are not always observed. \textit{Journal of the American Statistical Association}, \textbf{89}:846--866.
	
	\item Rubin, D. B. (1976). Inference and missing data. \textit{Biometrika}, \textbf{63}:581-–592.
	
%
	\item Sautory, O. (2003). CALMAR2: A new version of the
	CALMAR calibration adjustment program. \textit{Proceedings of
	Statistics Canadas Symposium}. Available at: http://www.statcan.ca/english/freepub/11-522-
		XIE/2003001/session13/sautory.pdf.
	
	\item Scharfstein, D. O., Rotnitzky, A., and Robins, J. M. (1999). Adjusting for nonignorable drop-out using semiparametric nonresponse models (with discussion and rejoinder). \textit{Journal of the American Statistical Association}, \textbf{94}:1096--1120.
\end{description}
\newpage
\appendix
\section*{Appendix}
\subsection*{Proof of Equation (\ref{Taylor_exp})}
We start by noting that $\widehat{t}_{MR}$ involves $J+L+3$ estimators: $\widehat{\ALPHA}=(\widehat{\ALPHA}^1, \dots, \widehat{\ALPHA}^J)^\top$, $\widehat{\BETA}=(\widehat{\BETA}^1, \dots, \widehat{\BETA}^L)^\top$, $\widehat{\ETA}_p$, $\widehat{\ETA}_m$ et $\widehat{\TAU}$. For this reason, we write $\widehat{t}_{MR} \equiv \widehat{t}_{MR}(\widehat{\ALPHA}, \widehat{\BETA}, \widehat{\ETA}_p, \widehat{\ETA}_m, \widehat{\TAU}).$  These estimated parameters are obtained by solving the following estimating equations:
\begin{align*}
\widehat{S}_{\bm{\alpha}}^{(j)}(\bm{\alpha}^{(j)}) &= \sum_{i \in S}w_i \frac{r_i-p^{(j)}(\mathbf{v}_i^{(j)}, \bm{\alpha}^{(j)})}{p^{(j)}(\mathbf{v}_i^{(j)}, \bm{\alpha}^{(j)})\{1-p^{(j)}(\mathbf{v}_i^{(j)}, \bm{\alpha}^{(j)})\}}\frac{\partial p^{(j)}(\mathbf{v}_i^{(j)}, \bm{\alpha}^{(j)})}{\partial \bm{\alpha}^{(j)}} =\mathbf{0}, \  j=1, \ldots, J; \\
\widehat{S}_{\bm{\beta}}^{(\ell)} (\bm{\beta}^{(\ell)}) &= \sum_{i \in S_r}w_i\{y_i-m^{(\ell)}(\mathbf{v}_i^{(\ell)}, \bm{\beta}^{(\ell)})\}\frac{\partial m^{(\ell)}(\mathbf{v}_i^{(\ell)}, \bm{\beta}^{(\ell)})}{\partial \bm{\beta}^{(\ell)}} = \mathbf{0},  \  \ell=1, \ldots, L; \\
\widehat{U}_p(\bm{\alpha}, \bm{\eta}_p) &= \sum_{i \in S}w_i(r_i-\mathbf{U}^\top_{p_i}\bm{\eta}_p)\mathbf{U}_{p_i} =\mathbf{0}; \\
\widehat{U}_m(\bm{\beta}, \bm{\eta}_m) &= \sum_{i \in S_r}w_i(y_i-\mathbf{U}^\top_{m_i}\bm{\eta}_m)\mathbf{U}_{m_i} =\mathbf{0}; \\
\widehat{U}_{\TAU}(\bm{\alpha}, \bm{\beta}, \bm{\eta}_p, \bm{\eta}_m, \bm{\tau}) &= \sum_{i \in S_r} w_i \frac{1-\widehat{p}_i}{\widehat{p}_i}(y_i-\mathbf{h}_i^\top\bm{\tau})\mathbf{h}_i = \mathbf{0}.
\end{align*}

Let $\ALPHA^\bullet$, $\BETA^\bullet$, $\ETA_p^\bullet$, $\ETA_m^\bullet$ and $\TAU^\bullet$ denote the probability limits of  $\widehat{\ALPHA}$, $\widehat{\BETA}$, $\widehat{\ETA}_p$, $\widehat{\ETA}_m$ and $\widehat{\TAU}$. Let $\mathbf{h}_i^{\bullet \top}=(1, m_i^\bullet)^\top,$ where $ m_i^\bullet$ is the probability limit of $\widehat{m}_i$ and $p_i^\bullet$ is the probability limit of $\widehat{p}_i.$ In the sequel, for ease of notation, we write $\widehat{\mathbf{S}}_{\ALPHA}^{(j)\bullet}$ for $\widehat{\mathbf{S}}_{\ALPHA}({\ALPHA}^{(j)\bullet}),$ $\widehat{\mathbf{S}}_{\BETA}^{(j)\bullet}$ for $\widehat{\mathbf{S}}_{\BETA}({\BETA}^{(j)\bullet}),$ $\widehat{U}_p^\bullet$ for $\widehat{U}_p(\bm{\alpha}^{\bullet}, \bm{\eta}_p^{\bullet}),$ $\widehat{U}_m^\bullet$  for $\widehat{U}_m(\bm{\beta}^{\bullet}, \bm{\eta}_m^{\bullet}),$  $\widehat{U}_{\TAU}^\bullet$ for $\widehat{U}_{\TAU}(\bm{\alpha}^{\bullet}, \bm{\beta}^{\bullet}, \bm{\eta}_p^{\bullet}, \bm{\eta}_m^{\bullet}, \bm{\tau}^{\bullet}),$ and $\widehat{t}_{MR}^\bullet$  for  $\widehat{t}_{MR}(\ALPHA^\bullet, \BETA^\bullet, \ETA_p^\bullet, \ETA_m^\bullet, \TAU^\bullet)$. \\

Using a first-order Taylor expansion, we first write:
\begin{align*}
&(i)  &\widehat{\ALPHA}-\ALPHA^\bullet &= -\E^{-1}\left(\frac{\partial \widehat{\mathbf{S}}_{\ALPHA}^\bullet}{\partial\ALPHA}\right)\widehat{\mathbf{S}}_{\ALPHA}^{\bullet}+o_p(n^{-1/2}); \\
&(ii)  &\widehat{\BETA}-\BETA^\bullet &= -\E^{-1}\left(\frac{\partial \widehat{\mathbf{S}}_{\BETA}^\bullet}{\partial \BETA}\right)\widehat{\mathbf{S}}_{\BETA}^{\bullet}+o_p(n^{-1/2}); \\
&(iii)  &\widehat{\ETA}_p-\ETA_p^\bullet &= -\E^{-1}\left(\frac{\partial \widehat{U}_p^\bullet}{\partial \ETA_p}\right) \widehat{U}_p^\bullet + \E^{-1}\left(\frac{\partial \widehat{U}_p^\bullet}{\partial \ETA_p}\right)\E\left(\frac{\partial \widehat{U}_p^\bullet}{\partial \ALPHA}\right)\E^{-1}\left(\frac{\partial \widehat{\mathbf{S}}_{\ALPHA}^\bullet}{\partial\ALPHA}\right)\widehat{\mathbf{S}}_{\ALPHA}^{\bullet}+o_p(n^{-1/2}); \\
&(iv)  &\widehat{\ETA}_m-\ETA_m^\bullet &= -\E^{-1}\left(\frac{\partial \widehat{U}_m^\bullet}{\partial \ETA_m}\right) \widehat{U}_m^\bullet + \E^{-1}\left(\frac{\partial \widehat{U}_m^\bullet}{\partial \ETA_m}\right)\E\left(\frac{\partial \widehat{U}_m^\bullet}{\partial \BETA}\right)\E^{-1}\left(\frac{\partial \widehat{\mathbf{S}}_{\BETA}^\bullet}{\partial \BETA}\right)\widehat{\mathbf{S}}_{\BETA}^{\bullet}+o_p(n^{-1/2}); \\
&(v) &\widehat{\TAU}-\TAU^\bullet &= -\E^{-1}\left(\frac{\partial \widehat{U}_{\TAU}^\bullet}{\partial \TAU}\right)\widehat{U}_{\TAU}^\bullet  -\E^{-1}\left(\frac{\partial \widehat{U}_{\TAU}^\bullet}{\partial \TAU}\right)\E\left(\frac{\partial \widehat{U}_{\TAU}^\bullet}{\partial \ALPHA}\right)(\widehat{\ALPHA}-\ALPHA^\bullet) \\ &&&\qquad -\E^{-1}\left(\frac{\partial \widehat{U}_{\TAU}^\bullet}{\partial \TAU}\right)\E\left(\frac{\partial \widehat{U}_{\TAU}^\bullet}{\partial \BETA}\right)(\widehat{\BETA}-\BETA^\bullet) \\ &&&\qquad  -\E^{-1}\left(\frac{\partial \widehat{U}_{\TAU}^\bullet}{\partial \TAU}\right)\E\left(\frac{\partial \widehat{U}_{\TAU}^\bullet}{\partial \ETA_p}\right)(\widehat{\ETA}_p-\ETA_p^\bullet)  \\ &&&\qquad -\E^{-1}\left(\frac{\partial \widehat{U}_{\TAU}^\bullet}{\partial \TAU}\right) \E\left(\frac{\partial \widehat{U}_{\TAU}^\bullet}{\partial \ETA_m}\right)(\widehat{\ETA}_m-\ETA_m^\bullet)+o_p(n^{-1/2}).
\end{align*}
Also, we have
\begin{align*}
\widehat{t}_{MR}(\widehat{\ALPHA}, \widehat{\BETA}, \widehat{\ETA}_p, \widehat{\ETA}_m, \widehat{\TAU}) &=\sum_{i \in S} w_i\frac{r_i}{\widehat{p}_i}y_i+\sum_{i \in S} w_i\left(1-\frac{r_i}{\widehat{p}_i}\right)\mathbf{h}_i^\top\widehat{\TAU} \nonumber \\
&= \sum_{i \in S} w_i\frac{r_i}{p_i^\bullet}y_i+\sum_{i \in S}w_i\left(1-\frac{r_i}{p_i^\bullet}\right)\mathbf{h}_i^{\bullet \top}\TAU^\bullet \nonumber \\ &\qquad +\E\left(\frac{\partial \widehat{t}_{MR}^\bullet}{\partial \ALPHA}\right)(\widehat{\ALPHA}-\ALPHA^\bullet)  + \E\left(\frac{\partial \widehat{t}_{MR}^\bullet}{\partial \BETA}\right)(\widehat{\BETA}-\BETA^\bullet) \nonumber \\ &\qquad + \E\left(\frac{\partial \widehat{t}_{MR}^\bullet}{\partial \ETA_p}\right)(\widehat{\ETA}_p-\ETA_p^\bullet) + \E\left(\frac{\partial \widehat{t}_{MR}^\bullet}{\partial \ETA_m}\right)(\widehat{\ETA}_m-\ETA_m^\bullet) \nonumber \\ &\qquad + \E\left(\frac{\partial \widehat{t}_{MR}^\bullet}{\partial \TAU}\right)(\widehat{\TAU}-\TAU^\bullet) + o_p\left(\frac{N}{\sqrt{n}}\right).
\end{align*}
After some algebra, we obtain
\begin{equation*}
\widehat{t}_{MR} = \sum_{k \in S} w_k\psi_k+o_p\left(\frac{N}{\sqrt{n}}\right),
\end{equation*} 
where
\begin{align*}
\psi_k &=y_k-\left(1-\frac{r_k}{p_k^\bullet}\right)(y_k-\mathbf{h}_k^{\bullet \top}\TAU^\bullet)\nonumber \\
&+ \sum_{j=1}^J\mathbf{A}_{\boldsymbol \alpha}^{(j)\bullet}\frac{r_k-p^{(j)}(\mathbf{v}_k, {\ALPHA}^{(j)\bullet})}{p^{(j)}(\mathbf{v}_k, {\ALPHA}^{(j)\bullet})\left(1-p^{(j)}(\mathbf{v}_k, {\ALPHA}^{(j)\bullet})\right)}\frac{\partial p^{(j)}(\mathbf{v}_k, {\ALPHA}^{(j)\bullet})}{\partial \ALPHA^{(j)}}\nonumber\\
&+ \sum_{\ell=1}^L\mathbf{A}_{\boldsymbol \beta}^{(\ell)\bullet}r_k\left(y_k- m^{(\ell)}(\mathbf{v}_k, {\BETA}^{(\ell)\bullet})\right)\frac{\partial m^{(\ell)}(\mathbf{v}_k, {\BETA}^{(\ell)\bullet})}{\partial \BETA^{(\ell)}}\nonumber \\
&+ \mathbf{A}_{p}^{\bullet}(r_k-\mathbf{U}_{pk}^{\top})\mathbf{U}_{pk}+\mathbf{A}_{m}^{\bullet}r_k(y_k-\mathbf{U}_{mk}^{\top})\mathbf{U}_{mk}\nonumber\\
&+\mathbf{A}_{\TAU}^{\bullet}r_k\frac{1-p^{(j)}(\mathbf{v}_k, {\ALPHA}^{(j)\bullet})}{p^{(j)}(\mathbf{v}_k, {\ALPHA}^{(j)\bullet})}(y_k-\mathbf{h}_k^{\bullet \top}\TAU^{\bullet})
\end{align*}
with 
\begin{align*}
\mathbf{A}_{\boldsymbol \alpha}^{(j)\bullet}=-\E\left(\frac{\partial \widehat{t}_{MR}^\bullet}{\partial \ALPHA^{(j)}}\right)\E^{-1}\left(\frac{\partial \widehat{\mathbf{S}}^{\bullet}_{{\ALPHA}^{(j)}}}{\partial\ALPHA^{(j)}}\right)+\E\left(\frac{\partial \widehat{t}_{MR}^\bullet}{\partial\ETA_p}\right)\E^{-1} \left(\frac{\partial \widehat{U}^{\bullet}_{{p}}}{\partial \ETA_p}\right)\E\left(\frac{\partial \widehat{U}^{(j)\bullet}_{{p}}}{\partial \ALPHA^{(j)}}\right)\E^{-1}\left(\frac{\partial \widehat{\mathbf{S}}^{\bullet}_{{\ALPHA}^{(j)}}}{\partial\ALPHA^{(j)}}\right), 
\end{align*}
\begin{align*}
\mathbf{A}_{\boldsymbol \beta}^{(\ell)\bullet}=-\E\left(\frac{\partial \widehat{t}_{MR}^\bullet}{\partial \BETA^{(\ell)}}\right)\E^{-1}\left(\frac{\partial \widehat{\mathbf{S}}^{\bullet}_{{\BETA}^{(\ell)}}}{\partial\BETA^{(\ell)}}\right)+\E\left(\frac{\partial \widehat{t}_{MR}^\bullet}{\partial \ETA_m}\right)\E^{-1} \left(\frac{\partial \widehat{U}^{\bullet}_{{m}}}{\partial \ETA_m}\right)\E\left(\frac{\partial \widehat{U}^{(\ell)\bullet}_{{m}}}{\partial \BETA^{(\ell)}}\right)\E^{-1}\left(\frac{\partial \widehat{\mathbf{S}}^{\bullet}_{{\BETA}^{(\ell)}}}{\partial\BETA^{(\ell)}}\right), 
\end{align*}

\begin{align*}
\mathbf{A}_{p}^{\bullet}=-\E\left(\frac{\partial \widehat{t}_{MR}^\bullet}{\partial\ETA_p}\right)\E^{-1}\left(\frac{\partial \widehat{U}^{\bullet}_{{p}}}{\partial \ETA_p}\right) ,
\end{align*}

\begin{align*}
\mathbf{A}_{m}^{\bullet}=-\E\left(\frac{\partial \widehat{t}_{MR}^\bullet}{\partial\ETA_m}\right)\E^{-1}\left(\frac{\partial \widehat{U}^{\bullet}_{{m}}}{\partial \ETA_m}\right),
\end{align*}
and
\begin{align*}
\mathbf{A}_{\TAU}^{\bullet}&=-\E^{-1}\left(\frac{\partial \widehat{U}^{\bullet}_{{\TAU}}}{\partial \TAU}\right)\widehat{U}^{\bullet}_{{\TAU}}  -\E^{-1}\left(\frac{\partial \widehat{U}^{\bullet}_{{\TAU}}}{\partial \TAU}\right)\E\left(\frac{\partial \widehat{U}^{\bullet}_{{\TAU}}}{\partial \ALPHA}\right){\mathbf{A}}^{\bullet}_{\ALPHA} -\E^{-1} \left(\frac{\partial \widehat{U}^{\bullet}_{{\TAU}}}{\partial \TAU}\right)\E\left(\frac{\partial \widehat{U}^{\bullet}_{{\TAU}}}{\partial \BETA}\right){\mathbf{A}}^{\bullet}_{\BETA} \nonumber \\ &\qquad  -\E^{-1}\left(\frac{\partial \widehat{U}^{\bullet}_{{\TAU}}}{\partial \TAU}\right)\E\left(\frac{\partial \widehat{U}^{\bullet}_{{\TAU}}}{\partial \ETA_p}\right){\mathbf{A}}^{\bullet}_{\ETA_p} -\E^{-1}\left(\frac{\partial \widehat{U}_{{\TAU}}^\bullet}{\partial \TAU}\right) \E\left(\frac{\partial \widehat{U}^{\bullet}_{{\TAU}}}{\partial \ETA_m}\right){\mathbf{A}}^{\bullet}_{\ETA_m}.
\end{align*}
\end{document}